\documentclass[12pt,preprint]{aastex}

\shorttitle{DUST IN THE PHOTOSPHERIC ENVIRONMENT }
\shortauthors{T. Tsuji}

\usepackage{aastexug}
\usepackage{natbib}
\begin{document}

\title{DUST IN THE PHOTOSPHERIC ENVIRONMENT: UNIFIED CLOUDY MODELS OF 
 M, L, AND T DWARFS }

\author{TAKASHI TSUJI}
\affil{Institute of Astronomy, The University of Tokyo, \\
Mitaka, Tokyo, 181-0015, Japan}
\email{ttsuji@ioa.s.u-tokyo.ac.jp}

\begin{abstract}

We report an attempt of constructing unified cloudy models for M, L, 
and T dwarfs. For this purpose, we first discuss opacities 
as well as thermochemical properties of the cool and dense matter.
Below about 2000\,K, refractory material condenses and dust 
will  play a major role as a source of opacity. 
Then a major problem in modeling the photospheres of very cool dwarfs
is how to treat dust, and especially  how dust could be
sustained in the static photosphere for a long time.
Under the high density of the photospheres of cool dwarfs, 
dust forms easily at the condensation temperature, $T_{\rm cond} $,
but the dust will soon grow larger than  its critical radius $ r_{\rm cr} $ 
(at which the Gibbs free-energy of condensation attains the maximum)
at the critical temperature $T_{\rm cr}$. 
Such large dust grains with $ r_{\rm gr} \gtrsim r_{\rm cr} $
will soon segregate from the gas and  precipitate below the photosphere.
For this reason,  dust  exists effectively only in the limited  
region of $ T_{\rm cr} \lesssim T \lesssim T_{\rm cond} $ in the photosphere,
and this means that a dust  cloud is formed deep in the 
photosphere rather than in the cooler surface region. 
With  this simple model of dust cloud, we show that the
non-grey model photosphere in radiative-convective equilibrium  
can be extended to $T_{\rm eff}$ as low as 800\,K. 
Since $  T_{\rm cond} \approx 2000$\,K for the first condensates such as
corundum and iron, the dust cloud is rather warm and necessarily located 
deeper in the photosphere ($\tau > 1$) for the cooler objects (note that
$T \approx T_{\rm eff}$ at $\tau \approx 1$). 
This explains why dust apparently shows little observable effect in 
T dwarfs.  For  warmer objects, the dust cloud which is always formed
at the same temperature range of $ T_{\rm cr} \lesssim T \lesssim 
T_{\rm cond} $  can be located nearer 
the surface ($\tau < 1$) and, for this reason, L dwarfs appear to be dusty. 
We show that the recently proposed spectral classification of L and T dwarfs
can consistently be interpreted by a single grid of our unified cloudy  
models with the thin dust cloud deep in the photosphere. 

\end{abstract}


\section{INTRODUCTION}

The present stellar spectral classification 
in terms of O, B, A, F, G, K, and M (with branching into R-N and S) has 
been used since the beginning of the 20\,th century.
For most of these spectral types, the modeling of stellar photospheres 
has now matured enough within the framework of the so-called classical 
theory of stellar photospheres in
general \citep[e.g.][]{kur94}. In the case of the latest M, S, and C types, 
however, the progress in modeling   photospheres  was rather slow
largely because of the extreme complexity of opacities  
dominated by molecules \citep[e.g.][]{gus94}. 
As for cool dwarfs, some initial attempts to include molecular
line opacities such as of H$_2$O were done a long time ago 
\citep[e.g.][]{aum69,tsu69}, and a systematic study of 
M dwarf model photospheres including several molecular opacity sources 
has been done by \citet{mou76} who provided a large grid for
M (sub) dwarfs for the first time.

More recent interest in cool dwarfs was motivated by
the progress of observations on faint cool dwarfs  on
one hand and by the increased interest on the so-called brown dwarfs
on the other. The possible presence of such  substellar objects 
was conceived a long time ago and their basic properties were predicted
already in the 1960's \citep[e.g.][]{hay63,kum63}.
Serious searches for brown dwarfs required accurate predictions
of observable properties and hence of the atmospheric structures
of brown dwarfs \citep[e.g.][]{burr93}. 
Some attempts to extend the model photospheres to the substellar regime 
with $T_{\rm eff}$ as low as 1000\,K have been done by \citet{sau94} 
for the case of zero metallicity and by \citet{tsu95a} for the 
solar as well as some sub-solar metallicities
in the same low temperature regime. 
These works revealed the importance of some new opacity sources such as 
H$_2$ collision-induced absorption (CIA) and methane (CH$_{4}$) 
under the extreme condition of the cool and dense photospheres. 
Also, many  model photospheres for a large parameter space were extended 
to $T_{\rm eff}$ as low as 1500\,K  by \citet{all95}  
and to the M dwarf regime by \citet{bre95}.

While gas phase chemistry has been sufficient in 
interpreting the classical spectral types from O to M, dust should be
the major ingredient in the very cool dwarfs. This possibility was
first noted on the late M dwarfs (Tsuji, Ohnaka, \& Aoki 1996a) and
confirmed on a larger sample of M dwarfs \citep{jon97}. 
However, the cool genuine  brown dwarf Gl\,229B finally discovered
by \citet{nak95} revealed no evidence for dust but showed strong bands 
of volatile molecules such as methane and water \citep{opp95}.
Such a result was quite consistent with the detailed thermochemical 
equilibrium calculations by \citet{feg96} who first showed that 
refractory elements including Ca, Al, Ti, V,
Mg, Si, Fe are removed by condensate cloud formation from the
observable photosphere and hence should not be observed.
In fact, the spectra of Gl\,229B  could rather be well interpreted by
the dust-free model developed  before the discovery of Gl\,229B 
\citep{tsu95a}, while brown dwarf candidate GD\,165B discovered by 
\citet{bec88} could first be explained by the dusty models 
\citep{tsu96b}. Models of Gl\,229B by other authors \citep[e.g.][]{
all96,mar96} also showed little effect of dust, and further
extended to the regime of giant planets \citep{burr97}.

Meanwhile progress in observations of  ultracool dwarfs  was
marvelous: Many new objects cooler than M dwarfs were discovered with the
DEep Near-Infrared Sky survey \citep[DENIS;][]{del97} as well as with the 
Two Micron All Sky Survey \citep[2MASS;][]{kir97} and the newly 
detected cool dwarfs were named as L dwarfs.
Cooler brown dwarf similar to Gl\,229B, named as T dwarf, was more 
difficult to find but several T dwarfs were finally  discovered with 
the 2MASS \citep{bur99,bur00a, bur00b} and with the Sloan Digital 
Sky Survey \citep[SDSS;][]{str99, tsv00}.
As is usually the case in the new field opened with the discovery of 
new objects, a great deal of effort has been done on how to classify the
newly defined L and T dwarfs. Initial attempts include  detailed 
classifications of L dwarfs  by \citet{kir99,kir00}
and by \citet{mart99} based on the optical spectra.
More recently, classification of L dwarfs was extended with the use of
the near infrared spectra \citep[e.g.][]{rei01,tes01}.
The first attempt on the detailed classification of T dwarfs was done 
by \citet{bur02}, and a unified classification scheme
for L and T dwarfs was proposed by \citet{geb02}.
Now, a major question is what is the exact meaning of the new spectral
classification in terms of L and T types.

Dust should certainly play a crucial role in interpreting the spectra 
of ultracool dwarfs, but a major problem
is how  dust could be sustained in the photosphere for a long time.
In fact, dust may easily segregate from the gas and precipitate below
the observable photosphere if dust grains grow larger, and 
Gl\,229B may represent such a case. On the other hand, late M dwarfs 
and GD\,165B may represent a case that dust grains can be sustained in 
the photosphere. Thus, we have considered two extreme cases to explain 
different types of ultracool dwarfs as reviewed elsewhere \citep{tsu00}; 
dusty model in which dust grains 
are sustained throughout the photosphere (case B) and dust-segregated
model in which all the dust grains have precipitated below the observable
photosphere (case C). Recently, more or less similar cases showing the
limiting effect of dust  were discussed by \citet{all01}. However, 
new observations  on a larger sample of L and T dwarfs revealed  
difficulty of the simple dusty and dust segregated models,
and  suggested a  more realistic model somewhere in between 
these two extreme cases \citep[e.g.][]{tin99}. 

It took sometime before we noticed that the extreme models, in which
the photosphere is filled with dust (case B) or fully depleted
of dust (case C), are physically unrealistic. We then considered dust 
formation and segregation in
a single model, the necessarily consequence of which  was the presence of 
a warm dust cloud deep in the photosphere.
This model was first applied  to a limited purpose of 
explaining the large optical flux depression observed in the  apparently
dust-free T dwarf Gl\,229B (Tsuji, Ohnaka, \& Aoki 1999). 
Although the optical flux depression itself may  be  explained 
by the strong alkali metal lines, as also noted by Burrows, Marley, \& 
Sharp (2000), the basic idea of the warm dust cloud deep in
the photosphere includes an important implication to be developed to a 
unified model photosphere of ultracool dwarfs \citep{tsu01}. We now 
examine such a possibility in detail in this paper. 

Recently, the dust cloud  model discussed in connection with the
planetary atmospheres has been extended to brown dwarfs
\citep[e.g.][]{lun89, ack01,mar02}.
Since brown dwarfs are just in between stars and planets,
approaches from both the stellar and planetary sides should be
useful and even complementary to each other. In the present paper,
however, we purposely restricted ourselves to the stellar approach and 
we hope to show how the methodology of
non-grey radiative-convective model stellar photosphere could be
extended to the photosphere of sub-stellar objects.

\section{ THERMOCHEMISTRY }

The most important basic input data, the composition of the chemical 
elements, is by no means well established yet even for the Sun
to which we referred (Sect.\,2.1). For model photospheres of ultracool 
dwarfs including brown dwarfs, the equation of
state (EOS) should be solved down to  $T \approx 500$\,K, and we
assume that  the ideal gas law can be applied up to the density 
regime of Kbar (1 Kbar $= 10^{9}$  dyn cm$^{-2} \approx 10^{3}$ atms) 
(Sect.\,2.2). Under the high density of the photospheres of cool dwarfs, 
the thermodynamical equilibrium can be well realized, and not only 
molecules but also dust grains are treated under the assumption
of the local thermodynamical equilibrium (LTE) (Sect.\,2.3).

\subsection{ Chemical Composition}

For ultracool dwarfs of the disk population, we may assume the chemical 
composition of the solar system, for which  the result by \citet{and89} 
is widely used. However, the solar system abundance 
is by no means well established yet and several revisions are needed
\citep[e.g.][]{lod98}. For example, the iron abundance should be represented 
by the meteoritic abundance, which was confirmed to be the same with
the solar photospheric abundance  at last \citep{bie91,hol91}.
Recently, a more drastic revision is proposed for the solar oxygen abundance:
log $A_{\rm O} = 8.69 \pm 0.05$ (Allende Prieto, Lambert, \& Asplund 2001)
on the scale of log $A_{\rm H} =12.0$ against the higher value of 
log $A_{\rm O} = 8.92$ \citep{and89}. The revised oxygen abundance is 
more consistent with the 
other disk population objects, but this problem should deserve further 
detailed analyses. Especially, the revised oxygen abundance and a slightly
updated carbon abundance of log $A_{\rm C} = 8.60$ \citep{gre91} are 
rather close and a difficult problem appears as will be discussed in 
Sect.7.3.  In the present work, we consider 34 elements of the solar 
system mixture summarized in Table 1, which is still based on \citet{and89} 
except for minor changes footnoted in the Table, but we must remember 
that the solar abundances are by no means well established yet.  
We also computed some models with the new (low) oxygen abundance, and the
results will be used to see the effect of oxygen abundance in Sect.7.3.

At the low temperatures and high densities realized in the photospheres of
ultracool dwarfs, most compounds  formed from H, C, N, O, and S are
volatile and the gas phase chemical equilibrium  plays a dominant
role. On the other hand, other abundant elements such as Mg, Fe, Si, and Al
condense in refractory compounds and phase transitions from gas to solid play
a major role. These 9 elements, together with the inert He, Ne, and 
Ar, constitute the most abundant elements in the solar system mixture 
and play a major role in determining the physical and chemical properties 
of the gaseous mixtures of the similar compositions.

\subsection{ Gas Phase Chemical Equilibrium }

After a survey of about 600 species in the chemical equilibria of 34 
elements \citep{tsu73}, we selected about 100 species in solving the 
chemical equilibrium during the iterations of modeling the photospheres. 
The thermochemical data used are mostly unchanged except for  
controversial cases such as  FeH (see Appendix), but some thermodynamical
data should certainly be updated as pointed out by \citet{lod02}.
For the purpose of the present work, however, we hope that the 
thermodynamical data of abundant molecules which are important as 
sources of opacity are relatively well established.

One important change from the chemical equilibrium familiar in stellar
photospheres  is that CO is no longer the major species of carbon and that 
this role is taken over by CH$_{4}$ at temperatures below  about 1000\,K 
in dense photospheres even for the case of $A_{\rm O} > A_{\rm C}$. A 
belief that all the carbon atoms are used in forming carbon monoxide (CO) 
in oxygen-rich photospheres is no longer true in the photospheres of 
substellar objects, and this is simply because CH$_{4}$ is thermodynamically 
more stable than CO at the very low temperature and high density. This fact
also implies that oxygen locked in CO is released and available to
form additional H$_{2}$O. For the same reason, almost entire oxygen is used 
to form H$_{2}$O rather than CO even in the carbon-rich gaseous mixture 
($A_{\rm O} < A_{\rm C}$) at the high density below about 1000\,K. This means 
that CH$_{4}$ and H$_{2}$O are the most abundant molecules independently of the
oxygen to carbon ratio, and there will be no spectral branching into M, S, and 
C types in the brown dwarf regime even if some brown dwarfs happen to
have chemical  peculiarity such as $A_{\rm O} < A_{\rm C}$. 

Other polyatomics such as  NH$_{3}$,  PH$_{3}$, H$_{2}$S are also abundant
and will play some role as opacity sources.
Although the gas phase chemical equilibrium in oxygen-rich photospheres is 
relatively simple even at very low temperature and 
high pressure, some interesting features are to be noted. For example,
a stable negative molecular ion  such as SH$^{-}$ appears as a sink of free 
electrons. It is to be noted, however, that the abundances of most volatile 
molecules remain unchanged by the formation of refractory condensates.
 
\subsection{ Condensation }

The phase transition from gas to solid in chemical equilibrium has been 
discussed  in the 1960's \citep[e.g][]{lor65,lar67} for the first time.
Since then, detailed studies of condensation have been developed in connection
with the primitive solar nebula \citep[e.g.][]{gro72} as well as with the 
giant planet atmospheres \citep[e.g.][]{feg94}. Also, the effect of 
condensation in molecular equilibrium was extended to the low pressure 
environments such as stellar envelopes \citep[e.g.][]{sha90}.
More recently, detailed analyses of the atmospheric chemistry including
condensation  have been done  with direct applications to the  brown dwarf 
atmospheres in mind \citep{burr99, lod99, lod02}.

Although several dozens of condensed species have been treated in the chemical
equilibrium by the works just referred to above, we found it not practical to 
include so many species in our modeling, which requires many iterative 
processes by itself. Also, it is difficult to specify dust opacity for
each individual species accurately at present. For these reasons,
we decided to represent the effect of dust formation by the three species,
namely corundum (Al$_2$O$_3$), iron (Fe), and enstatite (MgSiO$_3$).
In modeling photospheres, the first condensate should be 
most important since it determines the basic structure of the photosphere 
with  dust and gives a large effect on the  next condensates. It is known
that the first condensate in the solar composition mixture 
is ZrO$_2$ \citep{feg86,sha90},  which, however, is of low 
concentration  because of the low abundance of Zr. The next condensates are 
Al$_2$O$_3$ (at relatively low pressure) or iron (at relatively high pressure)
as can be confirmed in Fig.2 of \citet{lod99}. These condensates are quite 
abundant, and we regard corundum and iron as if they are the first condensates 
in our computation.
Also, we represent silicate minerals by MgSiO$_3$ which condenses at rather low
temperatures. Although forsterite (Mg$_{2}$SiO$_{4}$) condenses at the higher
temperature, we regard MgSiO$_3$ as the representative of the low temperature 
condensates. Since Fe, Mg, Si, and Al are the most abundant elements that 
produce refractory condensates, the total amount of condensed species 
can be approximated by these three species.  The resulting condensation lines 
of corundum, iron, and enstatite are graphically shown in Figs.2-4. These 
results agree well with those by a more detailed computation by \citet{lod99} 
(see her Fig.2) for a wide range of the total pressure. 

In our simplified treatment,
we first solved the gaseous equilibrium and then the chemical equilibrium
including condensation was  solved for Fe, Mg, Si, and Al. 
However, the condensation of silicates removes about 15\% of oxygen 
and will change the gaseous equilibrium as noted by \citet{lod02}.
We have confirmed that this effect is very important especially for
the case of the low oxygen abundance (log $A_{\rm O} = 8.69$), since the  
oxygen left in the gaseous phase is effectively log $A_{\rm O} \approx
8.69 - 0.07 \approx 8.62$ after 15\% oxygen is locked in silicate
and this is very close to the carbon abundance of log $A_{\rm C} = 8.60$.
In this case, the gaseous mixture is effectively close to the case of
$A_{\rm O}/A_{\rm C} \approx 1$, and H$_{2}$O abundance shows a drastic  
decrease while CH$_{4}$ abundance show an enhancement. Some details of 
this effect will be discussed in Sect.7.3.  
On the other hand, this effect  in the case of the high oxygen abundance
is just to reduce the oxygen-containing molecules by 0.07\,dex at the
largest and this effect will not be so serious for our present modeling
as will be shown in Sect.7.3.

As already noted by \citet{feg96}, refractory elements are removed 
from the gas phase mixture by the condensation.
Such a change in molecular abundances produces drastic effect on opacity if 
the non-volatile molecules are important sources of opacity. This is actually
the case of TiO, VO, FeH, and CaH, for which  we have included CaTiO$_3$
(perovskite), VO, Fe, and Ca$_2$MgSi$_2$O$_7$ (akermanite), respectively.
This is just to see the approximate effect of condensation on the molecular 
abundances, and these condensates (except for Fe) are not considered as 
sources of dust opacities for the reason outlined above.  It is to be noted, 
however, that Ca is the next abundant refractory elements after Al, and we 
are underestimating the effect of dust opacities.

Chemistry of a less abundant element is generally more complicated
because all the compounds composed of elements  more abundant than the 
element must be considered. This is the case of alkali metals
such as Na and K which suffer drastic depletions by NaAlSi$_{3}$O$_{8}$
and  KAlSi$_{3}$O$_{8}$, respectively, at true thermal equilibrium,
but may remain in atomic forms if highly refractive condensates
including Si and Al are removed from the photosphere, as noted by
\citet{lod99} and by \citet{burr01}. 
Also, detailed analysis of the  alkali metal chemistry by \citet{lod99}
showed that sulfides and chlorides should condense below about 1000\,K and
we considered the condensation of KCl and K$_{2}$S. We have used the 
thermochemical data for KCl and K$_{2}$S from the JANAF table \citep{cha85} 
and we applied the Gibbs free energy of formation by \citet{sha90} for 
other dust species discussed above. The thermodynamical data are still 
being up-dated and the recent revisions are reviewed by \citet{lod02}.

\section{ OPACITIES }

At the very  low temperature and high density environment to be expected 
in the photospheres of very low mass objects, new opacity problems
appear: First, some polyatomics such as CH$_{4}$ and  NH$_{3}$
must be added to the already known molecular line opacities (e.g. CO, 
H$_{2}$O, TiO, VO etc.) (Sect.3.1). Second,
collision-induced absorption (CIA) due to H$_{2}$--H$_{2}$ and H$_{2}$--He 
pairs plays an increasingly important role at higher densities (Sect.3.2). 
Third, as soon as various refractory 
condensates are formed as expected by the thermochemical
law, they act as efficient opacity sources over the entire spectral region
because of their large extinction cross-sections (Sect.3.3). Fourth,
atomic lines should not be neglected, since pressure broadened wings of 
strong alkali metal lines play a significant role as a source of opacities 
in cool and dense photospheres (Sect.3.4).

\subsection {Molecular Line Opacities}

We considered the effect of molecular line absorption by the band model method
during the construction of model photospheres, and used detailed linelist
to evaluate the final emergent spectra after the model has converged. 
In the band model method, molecular line opacity is characterized by 
two parameters, the straight mean absorption cross-section and the mean
line separation. Then, we apply  the Voigt-Analogue
Elsasser Band Model (VAEBM), which consists of an array of Voigt profiles of
equal intensity, spaced at equal intervals \citep{gol69}.  
For some cases (e.g. CH$_{4}$), however, we applied this method by its
simplest form  assuming that the line structure is completely smeared out
(also known as the Just Overlapping Line Approximation - JOLA) because
of the lack of the necessary molecular data. More details on the formulation 
of the band model method, together with its applicability and limitation, 
have been discussed elsewhere \citep{tsu84,tsu94}.
 
We have considered ro-vibration bands of diatomic (CO, OH, SiO) as well as
polyatomic (H$_{2}$O, H$_{2}$S, NH$_{3}$, PH$_{3}$, CH$_{4}) $ molecules,
together with the pure rotation transitions so far as they are allowed.
We have also considered electronic transitions of the refractory
molecules including TiO, VO, FeH, CaH, and MgH. The opacity data
used are summarized in the Appendix.

\subsection{ Collision-Induced Absorption (CIA) of H$_{2}$} 

At the very  high densities to be expected in the photospheres of 
ultracool dwarfs, absorption due to the dipole moment induced by 
collisions should be important for such abundant homo-nuclear molecules 
as H$_{2}$. In this case, individual lines are highly broadened 
because of the short time of the intermolecular interaction
inducing the dipole moment and hence completely smeared out. 
As a result, the collision-induced absorption (CIA) of H$_{2}$ dominates 
the whole infrared region by quasi-continuous absorption.
The importance of H$_{2}$ CIA in the photospheres of cool dwarf stars
has been recognized  at an early time \citep{lin69,tsu69}, 
but accurate cross-sections based on the detailed quantum mechanical 
analysis have been made available only recently by Borysow and her 
collaborators. The details are discussed for the case of  H$_{2}$ - 
He pairs \citep{bor89a, bor89b} as well as H$_{2}$ - H$_{2}$ pair 
\citep{bor90, zhe95}, and more recently by \citet{bor97}.
We have reproduced the absorption coefficients with the use of the computer
codes kindly made available by Dr. Borysow.

\subsection{ Dust Opacities }

For our present purpose, dust opacities are represented by the three species 
discussed in Sect.2.3, namely corundum (Al$_2$O$_3$), iron (Fe), and enstatite
(MgSiO$_3$). Further, we will show that only small dust grains can be 
sustained in the photosphere and hence will be important as sources of 
opacity (Sect.4.1). The mass absorption coefficient of dust grains of 
radius $r_{\rm gr}$ is 
   $$ \kappa ({\rm cm}^2/{\rm gram}) = 
   \pi r_{\rm gr}^2 Q_{\rm abs}/(4\pi r_{\rm gr}^3\rho/3), \eqno(1) $$
where $Q_{\rm abs}$ is the absorption efficiency factor and $\rho$ is the 
density of the grain. Then, $ \kappa $ depends little on the grain size so far
as $ r_{\rm gr} < 0.01\mu$m,  since 
   $$Q_{\rm abs} \propto r_{\rm gr}\eqno(2) $$ 
at $ x = 2 \pi r_{\rm gr}/\lambda << 1 $ (van de Hulst 1957).
For this reason, the resulting mass absorption coefficient remains unchanged 
for any size distribution so long as the grains are small enough, and dust 
opacities are  evaluated for $ r_{\rm gr} \approx 0.01\mu$m throughout.

For the grains of sub-micron size, dust opacities can be evaluated by 
a series expansion of the Mie formula \citep{van57}.
We used the optical constants for corundum by \citet{eri81}, who
presented the refractive index in the 0.4-2.0 $\mu$m wavelength range 
and the dielectric function in the 5-50 $\mu$m range. We have made
interpolation or extrapolation for the range where data are missing.
For iron, we used the experimental data by \citet{len66} and by 
\citet{ord88}. For enstatite, we applied the empirical opacity for 
warm oxygen-deficient circumstellar silicate by \citet{oss92}.
Although silicate abundance is represented by that of enstatite,
it actually represents all the silicates as noted in Sect.2.3 and
the opacity used also includes many silicates other than enstatite.

It is to be kept in mind that the actual dust opacity may be more
complicated. For example, dust opacities under astronomical
environment cannot be determined uniquely with the laboratory data, 
since dust grains in astronomical environment may never
consist of pure substance but may be composed of different species.
This fact may make the astronomical grains to be rather dirty
than clean  or to be heterogeneous (e.g. core-mantle structure) rather 
than homogeneous. Also, dust opacity depends on additional parameters 
such as  shape of the grains \citep[e.g.][]{ale94}. 
It is not possible, however, to consider all these complexities at present
and our dust opacities are very preliminary ones.

\subsection{ Alkali Metals }
  Non-refractory elements such as alkali metals remain in
mono-atomic gas at low temperatures near 1000\,K and their
strong resonance lines contribute significantly to suppress
the optical radiation  observed in cool brown dwarfs such as
Gl\,229B \citep{tsu99}. The exact evaluation of this rather
simple opacity source, however, appear to be difficult because of
the uncertainty in the line broadening theory of such strong lines
\citep{burr00}. In the present study, we assumed the classical
Lorentz line shape, but this should be replaced by a more
appropriate formula  when  it is available in the future.

\subsection{ Opacities per Gram of Stellar Material}
 
With the absorption cross-section for each species, we evaluate 
the absorption coefficient per gram of stellar material by solving 
chemical equilibrium for given temperature and gas pressure. 
Now, for each spectral mesh, we have the two parameters - the straight
mean absorption coefficient ${\kappa}_{i}$ and the mean line
separation ${d}_{i}$ for $i$-th species that is important as a line 
opacity source.  Then, the total absorption coefficient for molecular 
line opacity is simply the sum of $ {\kappa_{i}}$. The average line 
separation $d$ for the spectral mesh for all the contributing molecules 
($N$ species) can be estimated by
     $${ 1 \over d } = \{  \sum_{I=1}^{N} 
     { ({ {\kappa_i} \over {d_i}} )^{1/2} }  \}^2
     /{  \sum_{I=1}^{N} {\kappa_i} } . \eqno(3) $$
The results are the integrated mass absorption coefficient $\kappa$ and the 
average mean line separation $d$ for the given spectral mesh, chemical 
composition, temperature, and gas pressure. Then, the opacity distribution 
function (ODF) can be estimated within the framework of the VAEBM, by which
line broadenings due to turbulence and collision are considered at each
step of integration of the model photosphere. In this way, once the two 
parameters - ${\kappa}_{i}$ and  ${d}_{i}$  - for each molecule are prepared,
an approximate ODF can easily be estimated for any chemical composition, 
temperature, gas pressure, and micro-turbulent velocity, during the
iterative procedures in  construction of model photosphere.
An example of the extinction coefficients per gram of stellar material
for the chemical composition of Table 1, $T = 1008$\,K, and log $P_{\rm g}$
= 6.0 is shown in Fig. 1.  
The Rosseland and Planck mean opacities are also evaluated based on
these absorption coefficients \citep[e.g.][]{tsu95b}.

\section{ MODEL PHOTOSPHERES WITH DUST}

Unlike the case of cool giant stars where dust forms in the outflow, 
dust in cool dwarfs forms in the static photosphere and one problem 
is how dust could be sustained in the photosphere for a long time.
Thus, there should be some fundamental differences in the dust 
formation mechanisms in low and high luminosity stars.
Nevertheless, the basic processes of dust formation should
be understood by the same physical principle and we follow
a semi-empirical approach based on the homogeneous 
nucleation theory (Sect.4.1).  Then, it can be shown that
a dust layer (or may be referred to as a dust cloud) should be formed 
rather deep in the photosphere as a natural consequence of dust 
formation and segregation (Sect.4.2). We incorporate the dust cloud 
in the construction of model photosphere by the application of the 
classical  non-grey theory (Sect.4.3).

\subsection { Dust Formation and Segregation in the Photospheric
Environment }

The thermodynamical condition of condensation is well 
fulfilled in the photospheres of cool dwarfs (Sect.2.3). 
This is a necessary condition  but not the sufficient condition for 
condensation, since dust formed in the stellar photosphere is
usually in a form of  droplet which is subject to decay by
the surface tension force. For this reason, dust may form
when the super-saturation ratio $S = p/p_{\rm sat}  
(p_{\rm sat}$ is the saturation vapor pressure) exceeds unity,
but  will dissolve as soon as it is formed. The net effect is that 
effective nucleation does not start even if $S > 1$
and this phenomenon is generally referred to as super-saturation.

In the dense photosphere of cool dwarfs, however, we assume that 
the nucleation will eventually start sometime after the thermodynamical 
condition of condensation is fulfilled. According to the classical theory 
of homogeneous nucleation \citep[e.g.][]{has88}, the equilibrium 
concentrations of $n$-mer cluster $ C_e(n)$  and 
monomer $C_e(1)$ are related by
    $$  C_e(n) = C_1(1)exp[-\Delta G(n)/kT], \eqno(4) $$
where $\Delta G(n)$ is the Gibbs free energy of formation of $n$-mer 
given by
    $$ \Delta G(n) = -nkT {\rm In} S + 4\pi a_{0}^2 n^{2/3} \sigma, \eqno(5) $$
where $S$ is the super-saturation ratio, $a_{0}$ is a radius of monomer, 
and $\sigma$ is the surface tension of the condensate.
The Gibbs free energy of formation $\Delta G(n)$ shows a maximum at 
$n = n^{*}$ given by
    $$ n^{*} = {{8\pi a_{0}^2 \sigma^{3}} \over {3kT {\rm In}S} }. \eqno(6) $$
Since the Gibbs free energy cannot increase in any chemical process,
clusters with $ n < n^{*} $ cannot grow while those with $ n > n^{*} $ can.

This result implies that the cluster can grow stably only if it exceeds
the critical radius $r_{\rm cr}$ which corresponds to $  n^{*} $ given
by eqn.(6). In other words, stable dust grain can be formed only if 
the binding force which is proportional  to
the total number of monomers ($ \propto r_{\rm gr}^3$) exceeds the
destructive force due to surface tension which is proportional to
the surface area ($ \propto r_{\rm gr}^2$) at $ r_{\rm gr} = r_{\rm cr}$.
For this reason, dust has been regarded as formed when its size exceeds the
critical radius in general. For example,  in the case of dust formation in 
the mass-loss outflow from evolved stars,  it is indispensable that the
dust grows to be larger than the critical size in the dust forming 
region before it is ejected to rarefied circumstellar space,
since otherwise dust will resolve soon during the outflow and there is 
no chance for it to be formed again. 

In the case of stationary photosphere, however, clusters will grow 
irreversibly to the larger grains after the  cluster size exceeds the 
critical size. If dust grains are formed in the traditional sense, 
they will grow further to the more stable larger grains. But 
such larger grains may be difficult to be sustained in the static
photosphere and will finally segregate from the gaseous mixture.
Such large grains may no longer be  important as sources of opacity, since
they will fall below the visible photosphere or may be floating as 
impurities of a small filling factor in the photosphere. The very fact
that cool brown dwarfs such as Gl\,229B show little evidence for
dust could be explained this way  and this fact can be regarded as
evidence that dust grains have actually segregated from the ambient 
gaseous mixture in the photosphere of cool dwarfs \citep{tsu96b}.

On the other hand, if  dust grains remain to be smaller
than the critical size, they cannot grow to the stable solid particles.
This case that dust grains are unstable, however, is very interesting
since small dust grains may be destroyed easily but will soon be formed 
again  and {\it vice versa}, so long as the thermodynamical condition
of condensation is fulfilled (i.e.$~S > 1$). Thus, small dust grains whose 
abundance is determined by the chemical equilibrium always present 
in the static photosphere and  dust can be regarded as formed even if its 
size is below the critical size.
The reason why cluster sizes remain within the critical size cannot be
known exactly, but the clusters may anyhow be small enough just when
they are formed and also they may be destroyed by collisions to each
other at the high density of photospheres of cool dwarfs.

In the unstable regime of dust formation where thermal equilibrium is
realized, small dust grains are in detailed balance between formation 
and destruction, and this  gives a natural answer
why dust grains can be sustained in the static photosphere. In fact,
this will explain why some cool dwarfs including late M dwarfs appear to 
be dusty \citep[e.g.][]{jon97}, and further
why dust could survive as long as the lifetime of such long-lived stars
as M dwarfs.  Thus, somewhat paradoxically, 
a sufficient condition for the survival of dust in the photospheric
environment is that it is destroyed before it will grow too large.
Then, the notion of dust formation may be somewhat different in the
photospheric environment from that in circumstellar and interstellar
cases. We conclude that only small grains that failed to be the stable
large grains can be sustained in the static photosphere and
will play an important role in determining the photospheric structure.

\subsection{Unified Model Photospheres with the Dust Cloud} 

We now consider the photosphere in which the thermodynamical condition
of condensation is fulfilled. We assume that
dust  forms as soon as temperature is lower than the condensation 
temperature ($T_{\rm cond} $), but the dust will soon grow larger
than its critical radius $ r_{\rm cr} $   at the slightly lower temperature, 
say  $T_{\rm cr}$, which we referred to as the critical temperature.   
Thus,  in the region with  $T < T_{\rm cr}$ in the photosphere, dust will be 
large enough and may segregate from the gaseous mixture. 
Since it is  difficult to sustain such large dust grains in the photosphere,
they eventually precipitate below the photosphere where they may evaporate. 
Only in the region with $ T_{\rm cr} \la  T \la T_{\rm cond} $, the dust 
grains will be small enough ( $ r_{\rm gr} \la  r_{\rm cr} $) to be 
sustained in the photosphere and it is such small dust grains that play 
an important role as opacity sources. Thus, the dust effectively exists 
only in the restricted region located relatively deep in the photosphere, 
and this means that a dust layer (or a cloud) is formed in the photosphere. 
Thus, the cloud formation is a natural consequence of  considering not 
only dust formation but also its segregation.

Now, a major problem is to find the temperatures that define the  
dust cloud. The condensation temperature $ T_{\rm cond} $ is easily found
from the thermochemical computation  
to be $ T_{\rm cond} \approx 2000$\,K for  corundum and iron, which first 
form in the photospheres of ultracool dwarfs (as for detail, see Figs. 2-4). 
On the other hand, the critical temperature $ T_{\rm cr} $ is more difficult 
to find. This should in principle be determined from the detailed analysis 
of the dust-gas segregation process, but it still seems to be premature 
to solve this problem theoretically. Instead, we treat $ T_{\rm cr} $ 
as a free parameter to be found empirically. Since $ T_{\rm cr} $ 
essentially determines the thickness of the dust cloud, which in turn 
should give significant effect on observables, $ T_{\rm cr} $ could 
in principle be determined from observations.

So far, we have not yet specified  the critical radius 
$ r_{\rm cr}$ whose exact value is difficult to know, since it depends 
on the detail of the nucleation process. 
However,  the minimum value of astronomical grains known in the literature
is about $0.01 \mu$m, and this may imply that the grains of about this
size are already well stabilized. Then, we assumed that the critical size
$ r_{\rm cr} $ below which the grains are unstable and in 
detailed balance with the gaseous mixture is about $ 0.01 \mu$m 
(10 nanometer) or smaller. Then, the mass absorption coefficient of 
such dust grains, whose radii are sufficiently smaller than the wavelengths 
of interest, is almost independent of the grain size distribution 
(Sect.3.3), and our models are almost independent of the size distribution 
of dust grains under this assumption.

It is to be noted that the dust cloud can be found in all the cool 
dwarfs with $ T_{\rm eff} \la 2600$\,K. For objects with very low 
$ T_{\rm eff} $ near 1000\,K, the  dust cloud formed near the
dust condensation temperature ($ T_{\rm cond} \approx 2000$\,K) 
may be situated too deep in the photosphere (where $\tau_{\rm Ross} > 1$ , 
since $ T \approx T_{\rm eff}$ at $\tau_{\rm Ross} \approx  1$), while
the dust cloud may appear in the optically thin region of
the photosphere for relatively warm objects. This may explain
why L dwarfs appear to be dusty while cooler T dwarfs show
little evidence for dust. Thus our cloudy models will provide
unified explanation on the  dusty L dwarfs and apparently
dust-free T dwarfs by a single sequence of model photospheres.

Finally, our initial attempts considered three different cases
referred to as cases A (dust-free model), B (dusty model) and C 
(dust-segregated model) which correspond to $ r_{\rm gr} =0 $, 
$ r_{\rm gr} < r_{\rm cr}$ and $ r_{\rm gr} \ga r_{\rm cr}$, 
respectively, and  model photospheres were constructed based  on the
assumption that each case prevails throughout the photosphere \citep{tsu00}.
However, it is difficult to explain why different cases should apply to
different objects, while this difficulty is relaxed in our unified models.
In fact, such idealized cases will never be realized in nature and
the resulting models may be of little practical use for now (Sect.\,7). 
But they represent the extreme limiting cases of our unified 
cloudy models: The dusty model of case B  assumed that dust
formed at $T = T_{\rm cond}$  fills in the photosphere up to
the  surface, and thus represents a special extreme case of
$T_{\rm cr} = T_{0}$ where $T_{0}$ is the surface temperature. Also,
the dust-segregated  model of case C implies  that the dust grains
formed at $T = T_{\rm cond}$  segregate as soon as they are  formed, 
and this is another extreme case of $T_{\rm cr} = T_{\rm cond}$.

\subsection{Non-Grey Radiative-Convective Model  Photospheres} 

With the dust cloud outlined in the previous section,  model photospheres
based on the usual assumptions of hydrostatic and radiative equilibria are
constructed.  We assume LTE throughout. Also, plane parallel geometry can 
safely be assumed for the photospheres of cool dwarfs.
From the viewpoint of modeling the photosphere, the dust
cloud simply introduces a large increase of opacities at some layer
in the photosphere, and nothing is changed from the usual modeling
of stellar photosphere, even though  numerical computations are
somewhat more complicated. 

In the integration of model photosphere, we used the optical depth 
defined by the continuous opacity at 0.81$\mu$m and this independent 
variable is referred to as $\tau_{0}$.
The step of integration is either $\Delta$ log $\tau_{0} = 0.1$
($ T_{\rm eff} > 1400$\,K) or 0.05($ T_{\rm eff} \la 1400$\,K) for
log $\tau_{0} $  between -7.0 and 1.9.  We divided 
the spectral region between 0.25$\mu$m and 50$\mu$m into 
201 meshes during  iterations. We generally approximated the
line opacities by a three-step ODF in each spectral mesh, and thus any
quantity related to the radiation field is evaluated at  
201 $\times$ 3 frequency points. The final iteration, however, is done 
with 665 meshes (again with a three-step ODF). 

As is well known, convective energy transport is efficient at high 
densities of compact objects. We first constructed  pure radiative 
equilibrium model by our iterative temperature correction procedures. 
Then, if the model is found to be convectively unstable, we have 
applied  the local mixing length  theory (LMLT) to take into account 
the effect of convection. Then, we examine if the model satisfies
    $$  \pi F_{\rm rad}(\tau_{0}) + \pi F_{\rm conv}(\tau_{0}) = 
\sigma T_{\rm eff}^4, 
    \eqno(7) $$
where  $F_{\rm rad}(\tau_{0})$ is evaluated by solving the transfer equation 
and  $F_{\rm conv}(\tau_{0})$ is the convective flux given by the LMLT. If 
the model does not satisfy eqn.(7), we again applied the iterative 
temperature correction procedures. It is required, however, that the radiative 
flux should be $  F_{\rm rad}(\tau_{0}) = \sigma T_{\rm eff}^4/\pi 
- F_{\rm conv}(\tau_{0})$ rather than  $\sigma T_{\rm eff}^4/\pi$.
For the iterated  model, the LMLT  is again applied to evaluate convective 
flux, and these processes are repeated until convergence is obtained.

Here, however, at the onset of convection in the optically thin photosphere, 
say $ \tau_{0} \approx 0.01$, the convective flux of a few \% of the total 
flux appears suddenly, but it is difficult for radiative flux to decrease by 
the same amount suddenly because of the non-local character of the radiative 
flux. Thus it is difficult to satisfy eqn.(7) by better than a few \% at the
interface of the radiative and convetive zones in some cases. 
This problem is related to the inherent difficulty of the present LMLT, 
which is essentially a local theory but coupled with the 
non-local theory of radiative transfer. Probably, the flux error at the 
radiative-convective interface can be resolved by considering the 
overshooting of the convective cells to the radiative zone, and it may be 
of little meaning to try to reduce the flux error  of a few \% within 
the framework of the LMLT.

\section{PHYSICAL STRUCTURE OF THE UNIFIED CLOUDY MODELS}

We now apply the method of the non-grey model photosphere to the 
ultracool dwarfs with the dust cloud deep in the photosphere. 
The basic input parameters that specify a classical
model photosphere are chemical composition, effective temperature, 
surface gravity and micro-turbulent velocity. If dust forms in
the photosphere, however, we have to introduce an additional parameter, 
the critical temperature $T_{\rm cr}$, and we examine its effect by 
assuming four cases of $T_{\rm cr} = 1600, 1700, 1800  $ and 
1900\,K in addition to the extreme limiting cases of $T_{\rm cr} =
T_{0} $ ( case B) and $T_{\rm cr} =T_{\rm cond} $ ( case C) (Sect.5.1).
The chemical composition can be arbitrary changed but we restricted 
to the case of the solar system mixture discussed in Sect.2.1 in this 
paper. We also assumed log $g = 5.0$  throughout and the micro-turbulent 
velocity in the photosphere is assumed to be 1.0 km\,sec$^{-1}$, which 
is  near the solar value. We have computed grids of model photospheres 
for the six values of  $T_{\rm cr}$ noted above ( Sect.5.2). The presence 
of dust cloud results in a significant effect on the convective 
structure of cool dwarfs (Sect.5.3).

\subsection{Effect of the Critical Temperature}

We first examine the effect of the critical temperature 
$T_{\rm cr}$ on the thermal structure of the photosphere, 
and discuss its effect in some detail for the case of 
$ T_{\rm eff} = 1800$\,K  as an example. The resulting
thermal structures are shown in the top panel of Fig.2a for six values 
of $T_{\rm cr}$, namely $T_{\rm 0}$ (case B), 1600, 1700, 1800, 1900, and 
$T_{\rm cond}$ (case C). The lower boundary of the dust cloud is defined 
by the condensation line of corundum (Al$_2$O$_3$) or of iron (Fe) 
shown by the dot-dashed line in Fig.2a and the upper boundary by  
$T_{\rm cr}$ indicated by the plus sign for each model. Thus, the value 
of $T_{\rm cr}$ essentially specifies the thickness of the dust cloud. 
In Fig.2a, the resulting structures of our cloudy models are found between 
the dusty (cases B) and dust-segregated (case C) models as expected. 

The thermal structure of the case B model (i.e. $ T_{\rm cr} 
= T_{0}$) is rather peculiar in that it nearly coincides with the iron 
condensation line in the surface region. This is a general characteristic 
of the case B models whose photosphere is fully filled in by dust (see 
further Fig. 4) and this is because iron works as a thermostat: 
If iron grains are formed, they effectively absorb radiation and the
temperature of the iron forming region rises. Then the iron grains may 
evaporate and the temperature drops. Then, iron grains may form again
and so on. Thus, temperatures remain relatively high in the case B model 
throughout the photosphere. In another extreme model of case C  
( $ T_{\rm cr} = T_{\rm cond}$), the surface temperature is very low 
($ T_{0} \approx  800$\,K for $ T_{\rm eff} = 1800$ K). This is due to 
the cooling effect of the highly non-grey  opacities of the volatile 
molecules that are abundant in this case but no heating due to dust.

The formation of the dust cloud ($ T_{\rm cr} \la T \la T_{\rm cond}$, 
where $ T_{\rm cr} = 1600, 1700, 1800$ and 1900\,K with 
$ T_{\rm cond} \approx 2000$\,K )  results in a drastic change of
the thermal structure against the case of fully dusty model of case B.
Here, the role of iron as thermostat can still be seen in the cases of
$ T_{\rm cr} = 1600$ and  1700\,K, but it no longer plays a significant role
in the cases of $ T_{\rm cr} = 1800$ and  1900\,K  possibly because
the amount of iron grains may not be large enough for this effect to work.
On the other hand, dust has precipitated in the surface region above 
the dust cloud (where $ T <  T_{\rm cr}$) and volatile molecules dominate 
there. The photosphere is cooled appreciably by the cooling effect of the 
infrared  molecular bands of the volatile molecules as in our case C model.
The thermal structure of the upper photosphere is determined by
the balance of the heating due to dust grains in the cloud and the cooling by
the molecules above the cloud. The temperatures in the surface region
are lower for the higher $ T_{\rm cr}$, since the mass column density 
of the volatile molecules above the dust cloud should be larger. The 
photospheric structures of the cloudy models approach to that of the case 
B model in the region below the  dust cloud (where $ T >  T_{\rm cond}$).

In the case of  $ T_{\rm eff} = 1400$\,K  shown in Fig.2b, the thermal 
structure of the case B model  coincides with the iron condensation line 
only partly and shows considerable cooling in the surface region either 
by the infrared bands of silicate or by the molecules coexisting with 
the dust grains. The formation of the dust cloud results in a more 
drastic change of the thermal structure against the case of fully dusty 
model of case B than in the case of $ T_{\rm eff} = 1800$\,K (Fig.2a).
Here, the sudden increase of opacity due to the formation of dust cloud
at $ T = T_{\rm cr}$ results in very steep temperature gradient and, 
for this reason, temperature shows steep upturn, which starts at 
the shallower layer for the smaller $ T_{\rm cr}$. The photospheric 
structure of the cloudy model again approaches to that of the case B 
model in the region below the  dust cloud (where $ T >  T_{\rm cond}$).
On the other hand, dust-free region above the dust cloud (where 
$ T <  T_{\rm cr}$) is more extended and the cooling effect due to  
the infrared molecular bands of the volatile molecules is 
more effective than in the case of $ T_{\rm eff} = 1800$\,K (Fig.2a).
For this reason, the thermal structures of the cloudy models 
approach to that of the case C model in the surface region.

The case of the lower  $ T_{\rm eff}$ near  1000\,K, the effects of 
the dust cloud on the thermal structure  are essentially the same as 
in the case of $ T_{\rm eff} = 1400$\,K, except that its effect is  
less important on the structure of the surface region. For this reason, 
the structure of the cloudy model approaches to that of the case C in 
the upper photosphere, and to the case B model in the deeper region of 
the photosphere.  A preliminary version of the cloudy model
for the case of $ T_{\rm eff} = 1000$\,K with $ T_{\rm cr} = 1550$\,K 
was discussed before \citep{tsu99}.

In general,  iron (Fe) and corundum (Al$_{2}$O$_{3}$) condense at 
their condensation lines before the cloud forming region terminates at 
$T \approx T_{\rm cr}$. However, enstatite (MgSiO$_3$) is outside the 
cloud zone and this means that enstatite has segregated as soon as it 
is formed within the framework of our simplified treatment of the dust 
cloud. This may be possible because enstatite will easily form with 
corundum and/or iron as the seed nucleus and grow rapidly. For the same 
reason, other solid species that may form at the lower temperatures 
will precipitate as soon as they are formed. For this reason, only the 
dust species formed at relatively high temperatures above 
$T_{\rm cr} $ work as the active dust ( i.e. as sources
of opacity) and hence give significant effect on the photospheric structure.
This fact may simplify the construction of the cloudy models since it is 
enough to consider only a few high temperature condensates such as corundum 
and iron as sources of opacity. In some cases (e.g. $ T_{\rm eff} = 1400$\,K 
with $ T_{\rm cr} = 1600$\,K), however, a small amount of enstatite may form 
in the upper part of the dust cloud.

\subsection{Grids of the Unified Cloudy Models}

As shown in the previous subsection, the photospheric structure 
under the presence of the dust depends on a new parameter which we 
referred to as the critical temperature $ T_{\rm cr}$. We have computed
grids of model photospheres  with   $ T_{\rm cr} = 1600, 1700, 1800$ and
1900\,K for the effective temperature range between  800 and 2600\,K.  These 
grids are to be used to predict observables and to determine $ T_{\rm cr}$
by consulting with observed data (e.g. Sect.7.1). As an example,
we discuss the case of $ T_{\rm cr} = 1800$\,K and
the resulting models are shown in Fig.3, where the cloud zone is 
shown by the dotted area. 

In Fig.3, the locus of the Rosseland mean optical depth ($\tau_{R}$) 
unity is  shown by the dashed line. The locus of $\tau_{R} \approx 1$ 
roughly corresponds to $ T \approx  T_{\rm eff} $ 
but not exactly  because of the large  non-grayness of the opacity. 
Nevertheless, it is interesting to note that the dust cloud is 
situated in the regimes of  $\tau_{R} > 1$, $ \tau_{R} \approx 1$, and 
$ \tau_{R} < 1$ for the models of $T_{\rm eff} \la 1300, T_{\rm eff} 
\approx 1400 - 1600 $, and $T_{\rm eff} \ga 1700$\,K, respectively.
This result  indicates that the dust cloud should be situated within the 
optically thin regime  in the relatively warm objects with 
$ T_{\rm eff} \ga 1700$\,K, and this fact provides a natural
explanation why  L dwarfs appear to be dusty.
On the other hand, the dust cloud should be situated within the 
optically thick regime  in the relatively cold objects with 
$ T_{\rm eff} \la 1300$\,K, and this fact explains why  T dwarfs 
apparently show little evidence for dust. The cloud may be situated 
near the optical depth unity in the objects with $ T_{\rm eff} \approx 
1400 - 1600$\,K, which may just explain the early T dwarfs or L/T 
transition objects recently discovered by \citet{leg00}.

Finally, instead of presenting the grids for other $ T_{\rm cr} $'s,
we show the grids for the extreme limiting cases B and C by the
dashed and dotted lines, respectively, in Fig.4. It is to
be noted that all the case B models with $T_{\rm eff} \ga  1800$\,K
converges to the iron condensation line. 
All the cloudy models of a given $T_{\rm eff}$ can be found between
the models of cases B and C of the same $T_{\rm eff}$ as shown in Fig.\,2. 
Also all the cloudy models as well as  those by the cases B and C 
of a given $T_{\rm eff}$ degenerate to a single model for $T_{\rm eff} 
> 2600$\,K, in which dust no longer plays a major role. Then, our case C 
models represent the models of $T_{\rm eff} > 2600$\,K and  
extended to $T_{\rm eff} = 4000$\,K as shown by the solid lines in Fig.4.
It is noted that the models of $T_{\rm eff} = 2800 - 3200$\,K 
cross the dust condensation lines, but a small amount of
dust grains formed in the very surface of the photosphere gives
little effect on the physical properties of the models.

\subsection{Convection}

In the lower six panels of Figs.2a \& 2b, convective, radiative, and 
total fluxes normalized by $\sigma T_{\rm eff}^{4}/\pi$ are shown with
the dashed, dotted, and solid lines, respectively, for the six values of 
$ T_{\rm cr} $. In the case of $ T_{\rm eff} = 1800$\,K (Fig.2a),
the photosphere approaches to the wholly convective
in the deepest region in all the cases, and the structure of the
convective zone is rather simple. The cloudy models as well as
the dusty model of case B show essentially the same convective
structure as the case C model and the dust cloud gives
little effect on the convective structure. This may be because the
dust cloud is situated in the optically thin regime in the
photosphere and the amount of dust is not yet very large. 

The situation is quite different in the case of $ T_{\rm eff} = 1400$\,K 
(Fig.2b), in which the dust cloud is formed deep in the photosphere:
The temperature gradient is quite steep near the dust cloud because of 
the high  opacity due to the dust and the model is 
convectively unstable near $ T_{\rm cr}$. For this reason, our cloudy 
models show the outer convective zone in addition to the one
in the inner deep region. This new convective zone is
separated by an intermediate or detached radiative zone in all our
cloudy models, as discussed in the case of the hybrid model  of 
$ T_{\rm eff} = 1000$\,K with  $ T_{\rm cr} = 1550$\,K applied to Gl\,229B 
\citep{tsu99}. It is to be noted that such an outer convective zone 
does not appear in our case B model despite the large dust opacity. 
This may be because of the rather modest temperature gradient due to the 
heating effect of dust, and the large opacity does not necessarily 
induce convection.

In Fig.5a, some detail of the convective zone of the
cloudy models with $ T_{\rm cr} = 1800$\,K (Fig.3) are shown for
$ T_{\rm eff} $ between 800 and 2600\,K with a step of 200\,K.
In the models of $ T_{\rm eff} \ga 1800$\,K, the dust cloud
does not induce  convection as also noted on Fig.2a. 
In the model of $ T_{\rm eff} = 1600$\,K, the outer convective zone
separated from the inner convective zone appears, but the
convective flux is rather small. The outer convective zone develops
further as the dust cloud moves to the inner dense region in
the models with  $ T_{\rm eff} \la 1400$\,K.  The outer convective
zone, however, merges with the inner convective zone in the models 
with $ T_{\rm eff} \la 1000$\,K.

The convective velocities of the same models discussed above are shown 
in Fig.5b and  appear to be rather small in general.
For example, $V_{\rm conv} \approx 80$\,m sec$^{-1}$ in the hottest
model and as small as 10\,m sec$^{-1}$ in the coolest model. 
This is a natural consequence that only small convective
velocity is sufficient to transport the given energy flux
under the very high densities of the photospheres of ultracool dwarfs. 
Although the mixing length is 
assumed to be one pressure scale height throughout, the super-adiabaticity
is very small and the results depend little on the mixing length.

It is to be remembered that our models are based on the local mixing length 
theory (LMLT) and it is within the framework of the LMLT that a thin 
outer convective zone is predicted. However, it is unknown if the 
rather thin  convective zone can be stable or if it may induce some
dynamical effects. We defer more detailed analysis of the complicated 
structure of the convective zone in the cloudy model to future works, 
but the presence of the convectively unstable regime beside the one 
in the deeper region is a characteristic feature of the cloudy model 
of ultracool dwarfs. 

Even within the framework of the LMLT, however, it was rather difficult
to achieve the flux constancy within 1\% as is usually realized in
non-grey model photospheres without dust. As shown in  Fig.2 (also see 
Fig.5a), we had to allow flux errors of a few \% especially at the edges
of the outer convective zone. This should be due to essential difficulty
to match the convective flux determined locally and the radiative flux
of the non-local character (Sect.4.3). Also, appearance of the dust cloud
at $ T \approx T_{\rm cr}$ results in a sudden increase of opacity
and the requirement of the flux constancy had to be relaxed somewhat.

\section{PREDICTED EMERGENT SPECTRA }
 
Generally, stellar spectra could be interpreted in terms of
effective temperature $ T_{\rm eff} $, surface gravity $g$, 
chemical composition, and micro-turbulent velocity.
In ultracool dwarfs in which dust necessarily forms a cloud,
at least one additional parameter that specifies the nature of 
the cloud should be necessary and we represented it by the
critical temperature $ T_{\rm cr} $.
After a brief survey on the spectroscopic data (Sect.6.1), we
examine the effect of $ T_{\rm cr} $ (Sect.6.2) and $ T_{\rm eff} $
(Sect.6.3) on the emergent spectra, while other three parameters  
remain the same  as noted in Sect.5.

\subsection{Spectroscopic Data and Spectral Synthesis}

Once a model photosphere is at hand, the emergent spectra can readily
be computed for the given spectroscopic data. Unfortunately, however,
available spectroscopic data are far from satisfactory and this
is a major obstacle in predicting emergent spectra from
model photospheres of ultracool dwarfs. As an example, we discuss
the near infrared region for which spectroscopic as well 
observational data are relatively abundant.  

We have prepared a spectroscopic
database for the near infrared molecular lines including $^{12}$CO,
$^{13}$CO, C$^{17}$O \citep{gue83, cha83}, CN \citep{cer78, bau88}, OH 
\citep[GEISA;][]{jac99}, $^{28}$SiO, $^{29}$SiO, $^{30}$SiO \citep{lav81,tip81,
lan93}, and H$_{2}$O \citep[HITEMP;][]{rot97}. We have also tried other 
H$_{2}$O data  such as by \citet{par97} and found that the differences
with the HITEMP data are rather minor (typically, $ \Delta$log $ gf < 0.1$
and $\Delta\omega < 0.1$ cm$^{-1}$). 

For other molecules including CH$_{4}$, NH$_{3}$, PH$_{3}$, and 
H$_{2}$S, however, available linelists are 
limited to the low excitation lines of some restricted bands and are 
far short for our purpose. We regard the effect of these polyatomics as
pseudo-continua and apply the band model method to evaluate the
absorption coefficients (Sect.3.1). The electronic bands such as of
FeH, TiO, and VO have some effects near 1 $\mu$m, and we also treated
them as pseudo-continua. Also a few lines of K I \citep{wie69} are 
added but we do not intend to include comprehensive lilelists of atomic 
lines at present. 

The spectra were calculated with an interval of 0.05 - 0.1 cm$^{-1}$
and then convolved with the slit function which is assumed to be 
the Gaussian with FWHM typically 500\,km\, sec$^{-1}$ ($R = 600$).
The computation of the spectra for the cloudy models can be done with
any spectral synthesis code by simply adding the dust absorption and
scattering coefficients to the continuous absorption and scattering
coefficients, respectively, in the temperature range of 
$ T_{\rm cr} \lesssim T \lesssim T_{\rm cond} $.

\subsection{Effect of the Critical Temperature}

The effect of the dust cloud on the emergent spectra may 
be twofold: One is the direct effect due to the extinction by the dust
opacities and the other is the indirect effect due to the
change of the  structure by the dust cloud.
Also, since dust appears in the form of cloud, not only the thickness
of the cloud but also its location  as seen against optical
depth changes  even though the dust cloud always form
at the same temperature range of $T_{\rm cr} \la T \la T_{\rm cond}$. 
The spectra reflect all these effects. 

The predicted spectra between 1 and 4 $\mu$m based on the the
models of $T_{\rm eff} = 1800$\,K are shown for  $T_{\rm cr} =
1600, 1700, 1800$, and 1900\,K as well as for cases B and C in Fig.6a.
In these models of $T_{\rm eff} = 1800$\,K, the dust cloud is in the 
optically thin regime and the cloud itself may be still optically thin. 
Yet the dust cloud produces considerable extinction in the $J$ band 
region. Especially,  the dust cloud produces noticeable extinction 
in the case of $T_{\rm cr} = 1600$\,K, in which the spectrum is 
indistinguishable from that of case B. The strong H$_{2}$O 2.7 $\mu$m 
bands, which are sensitive to the thermal structure of the surface 
region, appear to be stronger for the higher $T_{\rm cr}$ for which 
the temperatures in the surface region are lower (Fig. 2a). 
It is interesting to see in Fig.6a that the $P$ and $R$ branches as well 
as the strong $Q$ branch of the CH$_{4}$ $\nu_{3}$ bands appears already 
at $T_{\rm eff} = 1800$\,K in the case of $T_{\rm cr} = 1800$\,K    
and strengthens towards higher $T_{\rm cr}$ (only $Q$ branch appears 
in the case of $T_{\rm cr} = 1700$\,K).

In the case of $T_{\rm eff} = 1400$\,K shown in Fig.6b, the predicted 
spectrum of case B is close to the black-body radiation of $T \approx 
1400$\,K, and this implies that the observable photosphere filled in 
by dust is now optically thick. Also, the difference of
the spectra for case B and for the cloudy models is quite large. 
This is  because the dust grains in the region of $T < T_{\rm cr}$
have precipitated in the cloudy models and the dust extinction
has decreased drastically. The optical thickness of the cloud in the 
observable photosphere is smaller for the higher $T_{\rm cr}$, and 
the extinction in the $J$ band region should decrease. At the same time, 
the column densities of volatile molecules above the cloud increase 
for the higher $T_{\rm cr}$ as the upper boundary of the dust cloud 
moves to the deeper layer and the molecular bands such as of H$_{2}$O 
and CH$_{4}$ strengthen.

Finally, in the case of $T_{\rm eff} = 1000$\,K shown in Fig.6c,
the predicted spectrum of case B shows weakly the $Q$ branch of the 
CH$_{4}$ $\nu_{3}$ bands, despite the large extinction by the dust. 
On the other hand, the
cloudy models show very strong bands of volatile molecules  such as 
of H$_{2}$O and CH$_{4}$, but the effect of dust is rather minor for
any value of $T_{\rm cr}$. This is because the dust cloud is now 
situated very deep in the photosphere below the observable layer 
and  volatile molecules dominate almost the entire region of the
observable photosphere above the cloud. This result implies that 
the presence of dust cloud deep in the photosphere gives little 
effect on the emergent spectra for very cool T dwarfs with 
$T_{\rm eff} \la 1000$\,K. This fact implies that rather simple 
dust-segregated models (case C) can be used for discussing observed 
spectra of cool T dwarfs, even though the dust deep in the photosphere  
gives considerable effect on the structure of the inner photosphere.

\subsection{Dependence on the Effective Temperature}

We show the effect of   $T_{\rm eff} $ on the emergent spectra
between 1 and 2.6 $\mu$m based on our unified cloudy models 
with  $T_{\rm cr} =1800$\,K in Fig.7 (we showed
the spectra by a step of 200\,K in $T_{\rm eff} $ to avoid
confusion although computations are done by a step of 100\,K).
We also included the models of $T_{\rm eff} > 2600$\,K in which 
dust no longer plays a major role and, for this purpose, we apply our
grid of the case C models. Thus, Fig.7 will give some idea on the 
change of the infrared spectra through M, L, and T dwarfs.   
We indicated  $T_{\rm eff} $ in units of 100\,K
on some spectra, and some spectral features are identified.  

For  $T_{\rm eff} > 2600$\,K, there is no effect of
dust, and dust extinction is still almost negligible
at $T_{\rm eff} \approx 2600$\,K even if a small amount of dust
is formed ($T_{\rm cr} < 1800$\,K). The effect of the dust extinction
gradually increases at the $J$ band region for the models with
$T_{\rm eff} < 2600$\,K so long as the dust cloud is located
within the optically thin regime. This effect is the largest  
at $T_{\rm eff} \approx 1600$\,K, where molecular bands appear to be
dimmed seriously. We confirmed that the spectra degenerate into
almost the same one for the models  of  $ T_{\rm eff} = 1500, 1600$, 
and 1700\,K at the $J$ band region, although this degeneracy disappears 
in the $H$ and $K$ band regions. This may be because
the dust cloud now penetrates into the optically thick regime
while the dust column density of the cloud itself may also increase
as the cloud moves to deeper (and hence denser) region. Because of these
two effects, the dust column density in the observable photosphere may 
remain nearly the same for $1500 \la T_{\rm eff} \la 1700$\,K. Finally, 
the dust extinction begins to decreases in $T_{\rm eff} \la 1400$\,K
according as the dust cloud moves to the deeper photosphere.
At the same time, abundances of the volatile molecules in the region 
above the cloud increase,  and  the molecular bands strengthen
rapidly towards cooler models.
   
The water bands  are visible at $T_{\rm eff} \approx 3000$\,K or higher
and and strengthen towards the lower  $T_{\rm eff}$'s. However, increase
of the water bands is rather modest in $ 1600 < T_{\rm eff} < 2000$\,K,
and this may be because the increased H$_{2}$O abundance with decreasing
$T_{\rm eff}$ is compensated for by the increased dust extinction.
In $T_{\rm eff} \la 1400$\,K, the increasing H$_{2}$O abundance and deceasing
dust in the observable photosphere with decreasing $T_{\rm eff}$
result in a rapid strengthening of the H$_{2}$O bands. The CH$_{4}$ 
2.2$\mu$m bands appear at $T_{\rm eff} \approx 1600$\,K by the low 
resolution ($R = 600$) employed in this figure while those at 1.6$\mu$m 
bands appear at  $T_{\rm eff} \approx 1500$\,K. At the same time, the
CO 2.3 $\mu$m bands, which are already prominent at $T_{\rm eff} \approx 
4000$\,K, almost disappear at $T_{\rm eff} < 1400$\,K. As we noted in
Sect.6.2, not only the $Q$ branch but also the $P$ and $R$ branches of the 
stronger CH$_{4}$ 3.3$\mu$m bands appear already at $T_{\rm eff} =1800$\,K 
for the cloudy models with $T_{\rm cr} = 1800$\,K or higher.

\section{COMPARISONS WITH OBSERVATIONS }

The emergent spectra of our cloudy models show characteristic
changes with $ T_{\rm cr} $ (Fig.6) as well with $ T_{\rm eff} $
(Fig.7). Such characteristics can most simply be represented by the
infrared colors, which in turn can be compared with observations.
It is shown that the effects of $ T_{\rm cr} $ and $ T_{\rm eff} $
on the infrared colors can be separated  and this fact makes it
possible to estimate $ T_{\rm cr} $ from the infrared colors (Sect.8.1).
The observed spectral features of ultracool dwarfs can directly be 
compared with the predicted ones based on our models (Sect.8.2). Also, 
some detail of water and methane bands will be discussed as a preliminary 
step towards detailed quantitative analyses of the spectra of ultracool 
dwarfs (Sect.8.3).

\subsection{Infrared Colors and  Estimation of $ T_{\rm cr} $ 
and $ T_{\rm eff} $ }

Infrared photometry in the $J, H,$ and $K$ (or $ K_{s}$) bands
has  extensively been applied to ultracool dwarfs including
L and T dwarfs. As examples, we show  observed infrared colors $J-K, J-H$,
and $H-K$ on the Mauna Kea Observatories (MKO) system
\citep{leg02} against the spectral types defined by 
\citet{geb02} in Fig.8a. An interesting feature is that all these 
infrared colors are not necessarily redder for the later types but 
show the red limit at about L5 - 6, as first noted by \citet{kir99}.

Based on the emergent spectra discussed in Sect.6, we tried to
predict the infrared colors on the $J, H, K$ photometry.
We apply the filter response function for the MKO system 
(Simon \& Tokunaga 2002; Tokunaga, Simon, \& Vacca 2002) 
and then the resulting colors are converted 
to the same scale as the observed colors by using the model  spectrum 
of Vega \citep{kur94}. The resulting predicted colors based on our 
grids of the models with $ T_{\rm cr} = 1600, 1700, 1800$, and 1900\,K 
as well as of our cases B and C models are shown in Fig.8b for comparison 
with observations. 

It is to be noted that all our cloudy models predict the presence of 
the red limits at about $ T_{\rm eff} \approx 1600$\,K while our cases B 
and C fail to do.  Thus, Fig.8 can be regarded as a clear demonstration 
that only the cloudy models are realistic enough in explaining the 
observed infrared colors. It is to be noted that the case B models 
predict much redder colors for cooler models while our case C models
never predict such red colors as observed for any value of $ T_{\rm eff}$.
This later result clearly demonstrates that the reddening of the
infrared colors in late M and L dwarfs is not due to the line blanketing 
effect of molecular bands but should mostly be due to the effect of
the dust.

The presence of the red limits in the infrared colors
should be a natural consequence of the presence of a dust cloud
deep in the photosphere. For a given value of $ T_{\rm cr}$, the
thickness of the cloud  always corresponds to $ T_{\rm cond} - T_{\rm cr}$ 
if seen by temperature and the cloud is situated at the same temperature
range of $ T_{\rm cr} \la T \la  T_{\rm cond}$, but the dust column 
density of the cloud increases for cooler $ T_{\rm eff}$, since
the cloud is situated in the inner (and hence denser) region if seen
by the optical depth. Thus, so far as the dust cloud is located
in the optically thin region, as is the case for  models with  
$ T_{\rm eff} \ga 1600$\,K (Fig.3), the infrared colors will be redder 
for cooler models because of the increased dust extinction due to
the increased dust column density. On the other hand, 
once the dust cloud penetrates into the optically
thick region in models with $ T_{\rm eff} \la 1500$\,K (Fig.3), 
the dust column density in the observable photosphere decreases
and, at the same time, the column density of the molecular gas 
above the cloud increases. For this reason,  the infrared colors 
turn to blueward until they finally merge with the infrared colors  
of the case C models at the coolest model (Fig.8b).    

The results in Fig.8b show that the infrared colors are redder
for the lower value of  $ T_{\rm cr}$ at a given  $ T_{\rm eff}$, and
this is because the thickness of the dust cloud is larger for the 
lower value of $ T_{\rm cr}$ at any given  $ T_{\rm eff}$.
Then, it is in principle possible to estimate  $ T_{\rm cr}$ by a
comparison of the observed (Fig.8a) and predicted (Fig.8b) colors. 
So far, we have assumed the same value of $ T_{\rm cr}$ for the range 
of  $ 800 \la T_{\rm eff} \la 2600$\,K, but this should most 
probably be unrealistic. For simplicity, however, we assume that  
late L dwarfs can be represented by the models of the same $ T_{\rm cr}$. 
Since the effect of the dust cloud on observables is more prominent 
in L dwarfs  than in T dwarfs, we are primarily interested in the 
nature of the cloud in L dwarfs. Then, we estimated that the maxima 
of the mean values of the observed $J-K, J-H$, and $H-K$, are about
1.8, 1.0, and 0.8, respectively, at about L5.5 (Fig.8a), and
these maxima can roughly be reproduced at about $ T_{\rm eff} = 1600$\,K
by the predicted  $J-K, J-H$, and $H-K$ with  $ T_{\rm cr} = 1800$ or
slightly lower (Fig.8b). From this result, we suggest that $ T_{\rm cr}$ 
should be close to 1800\,K and we apply  $ T_{\rm cr} = 1800$\,K in the 
rest of this paper. This result implies that the dust cloud may be rather 
thin at least in late L dwarfs.

Also, our preliminary analysis on the different observed data based on 
the $J, H,$ and $ K_{s}$ photometry \citep{kir99, kir00} arrived at the 
same conclusion of $ T_{\rm cr} = 1800$\,K \citep{tsu01}.
On the other hand, our previous suggestion of $ T_{\rm cr} = 1550$\,K for 
a model of  $ T_{\rm eff} = 1000$\,K was largely biased by the analysis 
of the  optical spectrum of Gl\,229B \citep{tsu99}.
However, we now know that the dust cloud is situated too deep
to give appreciable effect on the observed spectrum in the case of such 
a cool T dwarf, and the estimation of $ T_{\rm cr}$ should be
difficult in such a case. With this difficulty in mind, 
the modest dependence of the emergent spectra on  $ T_{\rm cr} $ for 
the case of 1000\,K (Fig.6c) can be used to constrain
somewhat the value of $ T_{\rm cr} $ and our revised analysis of
the optical spectrum of Gl 229B appeared to be consistent with 
$ T_{\rm cr} = 1800$\,K \citep{tsu01}.

It is to be noted, however, that the scattering in the observed data are 
rather large and this fact may suggest that the thickness of the dust 
cloud is not necessarily the same even at a given $ T_{\rm eff}$. 
In fact, the scattering of the observed colors (Fig.8a) can be explained 
by the cloudy models of  $ T_{\rm cr} \approx 1700 - 1850$\,K (Fig.8b).
As to the origin of the scattering, however, other effects such as of 
metallicity, age, mass, rotation, and other activities must also be 
considered. On the other hand, a safe conclusion is that the cases of 
$ T_{\rm cr} = 1600$ and 1900\,K, which predicts the infrared colors 
too red and too blue, respectively, compared 
with the observed ones, may be excluded. Also, an encouraging result 
is that all the three predicted colors show the red limits at 
$ T_{\rm eff} \approx 1600$\,K while the red limits of the three 
observed colors are all at about L5.5. This fact suggests that 
$ T_{\rm eff} \approx 1600$\,K at about L5.5, and this
result is almost free from the choice of $ T_{\rm cr}$.

Estimations of $ T_{\rm eff}$'s for other spectral types are more difficult. 
Colors are often used as indicators of stellar effective temperatures,
but accurate calibration of color against $ T_{\rm eff}$ is difficult 
for cool stars because of the severe line blanketing effect. 
As a preliminary attempt, we have estimated representative
colors for several spectral types from a curve drawn by free-hand on 
each color-spectral type plot of Fig.8a. Then, $ T_{\rm eff}$ is estimated 
for each color from Fig.8b, but the results depend on $ T_{\rm cr}$ for 
which we assumed 1800\,K for $J-K, J-H$ and $H-K$. 
The results are summarized in Table 2 and the mean  
$ T_{\rm eff}$ from the three colors is also given for each subtype.
The results from different colors are mostly within 100\,K of the mean 
$ T_{\rm eff}$ and thus internally consistent within about 100\,K.

\subsection{Spectral Indices}

As an example of the spectral indices employed in the spectral
classification, the H$_{2}$O 1.2 $\mu$m index
defined by \citet{geb02} is shown against their spectral type
in Fig.9a. For comparison, we calculated this index on our predicted 
spectra (such as shown in Fig.7 but transformed to the $F_{\lambda}$ scale) 
following the definition by \citet{geb02} and the result
is plotted on Fig.9b. The observed index shows  rather modest
increase in L dwarfs but shows  steep increase in T dwarfs.
On the other hand, the predicted index based on our cloudy models
shows  rapid upturn  at $ T_{\rm eff} \approx 1400$\,K which may correspond
to the early T types in agreement with observations. The predicted index
based on our case C models (dotted line in Fig.9b) shows  upturn
at about $ T_{\rm eff} \approx 1700$\,K while that based on our case B
models (dashed line) show no such rapid upturn in the T dwarf regime.

The observed CH$_{4}$ 2.2 $\mu$m index by \citet{geb02} is
shown in Fig.10a while the predicted one based on our models
in Fig.10b. The observed index shows that the methane 2.2 $\mu$m bands
appear at  early T types and our cloudy models predict that 
the bands appear at about  $ T_{\rm eff} \approx 1500$\,K in agreement 
with the results discussed in Sect.6.3. On the other hand, our case C 
models show that the CH$_{4}$ 2.2 $\mu$m bands appear  
at  $ T_{\rm eff} \approx 1700$\,K. Also, our case B models show only
weak methane bands in the T dwarf regime as expected.

As an example of atomic line features, the observed pseudo-EW's
of K\,I 1.2432\,$\mu$m given by \citet{bur02} are
plotted against their spectral type in Fig.11a. For comparison,
predicted EW's measured on the spectra based on our cloudy models 
(Fig. 7) are shown by the solid line while those based on our cases B 
and C models by the dashed and dotted lines, respectively, in Fig.11b. 
The observed features show  drop at late L types and our prediction 
based on the cloudy models also shows  rapid decrease at $ T_{\rm eff} 
\approx 1500$\,K, while this effect is not predicted by our case C models. 
The observed EW's again show  upturn in the T dwarf regime, and this 
effect is also predicted by our cloudy models. Although the rapid 
decrease of the K\,I EW's in late L dwarfs can be explaind by the 
case B models as well, the  K\,I line totally disappears below about 
$ T_{\rm eff} \approx 1500$\,K in the case B models. Thus the rather
complicated behaviour of the K\,I line can be explained only by our
cloudy models: The observed EW's 
of the K\,I line shows the minimum at about  L7, and this is because  
the dust column density and hence the dust extinction are the largest 
at about $ T_{\rm eff} \approx 1600$\,K in our cloudy models. 
According as the dust cloud moves to the inner region in the cooler
models, the dust extinction decreases and the K\,I line strengthens again.  
The increase of the K\,I EW, however, cannot be so large in T dwarfs 
even if the dust column density in the observable photosphere decreases, 
because of the unfavourable excitation of the K\,I line at the low 
temperatures (1.2432\,$\mu$m line originates from a level with 1.61 eV).

\subsection{Water and Methane Bands}

It is known that quantitative analyses of water bands showed serious 
difficulty in that predicted water bands appeared to be too strong 
compared with observations (e.g. Tinney, Mould, \& Reid 1993). 
In fact, this problem was one of the motivations that we proposed 
the presence of dust in late M dwarfs (Tsuji et al. 1996a). Another 
possibility often mentioned is the incompleteness of the present linelists 
\citep[e.g.][]{rot97, par97, vit97} which do not yet cover the high 
excited levels reasonably well. However, if the hot bands not yet included 
in the present linelist are added, predicted water bands will be stronger 
and the discrepancy with the observed ones may not be resolved.

Under such circumstances, we hoped that a solution can be provided by 
the recent revision
of the solar oxygen abundance \citep{alle01} noted in Sect.2.1, and we
discuss this effect in some details for the case of the L dwarf prototype, 
GD\,165B.  For this L3 $\pm$ 1 dwarf \citep{geb02}, we may assume $T_{\rm eff} 
\approx 1800$\,K (Table 2).  In Fig.12a, we compared the  spectrum observed 
by \citet{jon94} with the predicted spectra based on the cloudy model 
($ T_{\rm eff} = 1800$\,K and $ T_{\rm cr} = 1800$\,K) using the high oxygen 
abundance. The effect of silicate formation on H$_{2}$O abundance (Sect.2.3) 
is considered in the heavy solid line but  not considered in the thin line. 
The two cases show little difference on the scale of Fig.12a and the 
reduction of H$_{2}$O abundance by 15 \% (or -0.07 dex.) in the limited 
region of the photosphere gives only minor effect if  oxygen abundance is 
sufficiently high compared with the carbon abundance. 
 Also, it is to be noted in Fig.12a that the predicted water bands
appear to be stronger as compared with the observed ones.

In Fig.12b, we compared the  same observed spectrum  with the predicted 
spectra based on the cloudy model ($ T_{\rm eff} = 1800$\,K and 
$ T_{\rm cr} = 1800$\,K) using the low oxygen abundance. 
The reduction of oxygen from the high abundance (log $A_{\rm O}$ = 8.91) 
to the low abundance (log $A_{\rm O}$ = 8.69)  implies the reduction of  
$A_{\rm O} - A_{\rm C}$ and hence of the H$_{2}$O abundance by about a 
factor of 4 (with log $A_{\rm C}$= 8.60) so long as most carbon is in CO. 
For this reason, H$_{2}$O bands show rather large reduction for the
decrease of oxygen abundance by a factor of 2, even for the case in which
the effect of silicate formation on H$_{2}$O abundance is not considered
(the thin solid line). Further, if the effect of silicate formation on  
H$_{2}$O abundance is considered, the reduction of H$_{2}$O abundance cannot 
be 15\% but should be much larger, since the free gaseous oxygen is now very 
small (effectively equivalent to the case of log $A_{\rm O}$ - log $A_{\rm } 
\approx 0.02 $ or $A_{\rm O}/A_{\rm C} \approx 1.05 $ as noted in Sect.2.3). 
The result shown by the heavy solid line clearly shows drastic effect of
the silicate formation on H$_{2}$O abundance. Further, the CH$_{4}$
bands at 2.2\,$\mu$m are considerably enhanced. 

Thus, we confirmed that the reduction of the free oxygen due to the 
silicate formation noted by \citet{lod02} will give significant effect
especially if the low solar oxygen abundance \citep{alle01} is applied. 
However, this case is clearly not realistic, since water bands are too weak
while methane bands are too strong compared with the observation. Nevertheless 
this fact alone may not be a reason to reject the low oxygen abundance, 
since what matters is the relative abundance of oxygen to carbon.
In fact,  if the  solar oxygen abundance  had to be revised by as much as
a factor of 2, the solar abundances of carbon and nitrogen, which have been 
determined by more or less similar method as for oxygen,  should also be  
reexamined. Until such a problem can be resolved, we cannot apply the new 
solar oxygen abundance to L and T dwarfs.  For this reason, we  discussed 
the models based on the high oxygen abundance (Table 1) throughout this paper.

In contrast to the case of water, major difficulty in discussing 
methane bands is the lack of spectroscopic data that can be applied 
to temperatures above the room temperature, even though there are 
some data in HITRAN \citep{rot98} and GEISA \citep{jac99}. 
 Hopefully, such excellent databases should not be restricted to 
investigating Earth's  atmosphere but be extended for applications 
to stellar and substellar problems. Without having other possibilities 
at present, we applied 56702 lines of $^{12}$CH$_{4}$ available in 
the GEISA database to evaluate synthetic spectra based on our model 
of  $ T_{\rm eff} = 1000$\,K with  $ T_{\rm cr} = 1800$\,K, and the result 
is compared with the observed spectrum of Gl\,229B \citep{geb96, opp98, 
leg99} in Fig.13a. It is clear that the predicted spectrum largely
underestimates the strength of the observed methane bands, and this is
what was expected since the linelist does not include many lines from
the excited states.  Even the $\nu_{3}$  fundamental bands at 3.3 $\mu$m 
could not be predicted well except possibly for the strong $Q$ branch, 
and the situations with the 1.6 and 2.2 $\mu$m bands are even worse.
 
Then, we apply the band model opacity (Sect.3.1) to methane bands in 
its simplest form, that is as a smeared-out pseudo-continuum (but
we used the linelist for all the other molecules such as CO and 
H$_{2}$O).  The resulting predicted spectrum is compared with the 
observed one in Fig.13b, and the agreement is now quite
improved, especially for the 1.6 and 2.2 $\mu$m bands. For this reason,
we have used the band model opacity for methane throughout this paper,
and we hope that the predicted methane bands based on this simple
band model can be used for a while until a better linelist of methane
can be available in the future. It is unfortunate, however, that the 
opacity had to be tested by the observed data of brown dwarf while what 
we should do is the reverse, even though  the band model opacity 
of methane is tested by the linelist at least at the room temperature
(Fig.14 in the Appendix).

\section{ DISCUSSION }

\subsection{Dust Cloud in the Photospheric Environment}

A possibility that dust forms in stellar photosphere might have been 
conceived in the past but few attempts have been done to take the 
dust formation into account in modeling stellar photosphere.
Probably, a drawback of including dust in modeling stellar photosphere
may be because it was unknown how to treat dust in the photospheric 
condition. For example, it remained unknown how dust could be 
sustained in the static photosphere. We propose a solution 
to this problem that the dust grains smaller than the critical 
size are in detailed balance with the surrounding gaseous mixture
and hence can easily be sustained while grains larger than the
critical size grow larger and may not be sustained in the static
photosphere. The cloud formation is a natural consequence of the 
formation of dust at the condensation temperature $T_{\rm cond} $ 
and its segregation at the critical temperature $ T_{\rm cr}$, 
resulting in a dust cloud in the restricted region of 
$ T_{\rm cr} \la T \la T_{\rm cond} $.
Then,  the classical non-gray theory of the stellar photosphere
can directly be applied to the photosphere with a thin dust cloud
and no other ad-hoc assumption is introduced at all to the classical
theory of the stellar photosphere. Our basic assumption appears
to be confirmed  by the fact that the model photosphere based on this
assumption predicts the observed spectra and colors reasonably well 
(Sect.7).

Our model is not yet complete in that the critical temperature cannot 
be determined from the physical theory but had to be treated as a 
free parameter to be determined empirically.
Thus, our present model is semi-empirical in nature and we did
not  invoke the detail of the cloud formation mechanism. 
In this connection, an interesting attempt to determine the 
dust properties formed in the brown dwarfs was done 
by the analysis of the hydrodynamical processes induced by
the acoustic wave originating in the convective zone \citep{hel01}. 
We hope that such an analysis will be developed to
a fully {\it ab-initio} modeling including cloud formation.

Recently, formation of dust cloud has also been discussed by applying
the method of planetary atmospheres \citep{ack01, mar02}.
Although we agree in that the dust cloud may be formed
rather deep in the photosphere, there are some differences in detail.
For example,  they concluded that the grain sizes are as large as
10 - 100 $\mu$m while we assumed much smaller sizes below 0.01$\mu$m.
The observations of cool brown dwarfs such as Gl\,229B indicate
that the large dust grains cannot be sustained in the observable photosphere.
A question in the cloud model of \citet{mar02} is how such large grains 
can be sutained so nicely near the dust condensation temperature and 
why they do not fall further below that point. In our case, this 
problem is solved by considering only small grains that are in detailed 
balance with the gaseous mixture.  
Despite such a difference, the effect of the dust cloud on the thermal 
and convective structures was solved self-consistently both in their 
models and in ours. This has been a major issue in stellar model photosphere 
and this tradition is now extended to self-consistent models of
substellar objects, in radiative-convective equilibrium under the presence 
of the dust cloud, by Marley et al.(2002) and by our present work. We 
agree further that one free paremeter that specifies the thickness of the 
cloud had to be introduced; their $f_{rain}$ and our $T_{\rm cr}$. 
Thus, it appears that the cloud model based on the methodology of
the planetary atmosphere is also semi-empirical in nature.
This fact implies that it is still difficult to specify the nature of the 
cloud by the present theory of the planetary atmosphere as well as by 
that of the stellar photosphere.

\subsection{Physical Basis of the Spectral Classification of 
L and T Dwarfs}

We showed that a single grid of  model photospheres with a thin dust 
cloud explains the major spectral features as well as the infrared 
colors of L and T dwarfs consistently. Although some spectral indices 
such as of water may also be accounted for by the dust-segregated 
models of case C (Sect.7.2), the infrared colors can never be 
explained by the case B nor by the case C models (Sect.7.1).
This fact confirms that only the cloudy models provide the
physical basis for the spectral classification of L and T dwarfs. Then,
the systematic change of the colors and spectral features used in the 
spectral classification of L and T dwarfs can be interpreted
as the effect of temperature and this fact confirms that the proposed 
spectral classification of L and T dwarfs  can be a reasonable one 
in that it represents a temperature sequence, even though the spectral 
classification has been done on purely empirical basis following the 
tradition since the classical Harvard system \citep{kir99, kir00, 
geb02, bur02}.

It is to be noted, however, that the major spectral indices used in the
spectral classification of L and T dwarfs are not direct measures of stellar 
temperature as are usually the case in the spectral types from O to M, but are
largely controled by the presence of the dust cloud in the photosphere.
At the same time, L and T  types are not simply related to the amount of
dust  predicted by the chemical equilibrium theory while O - M types
are directly connected to the abundances of atoms, ions, and molecules 
predicted by the ionization and dissociation theory. What is more important 
in L and T dwarfs is the location of the dust cloud in the photosphere:
Since the dust cloud is always formed near the dust condensation temperature,
its relative location in the photosphere moves from deep ($\tau > 1$) to 
surface ($\tau > 1$) regions according as $T_{\rm eff}$ increases from  
T  to L dwarfs. It is this change of the location of the dust cloud that plays 
an imporant role in the characterization of the spectral types from L to T.  
Thus, although spectral types L and T can in principle  be interpreted
as a temperature sequence by the chemical thermodynamics, 
some non-equilibrium processes such as segregation, coagulation, and 
precipitation  of dust grains must be considered. At present, the effects of 
these complicated processes are represented by the formation of the thin dust 
cloud deep in the photosphere, and this simple model explains the
L -- T spectral sequence quite well. 

The spectral classification scheme of ultracool dwarfs, however, may not be
deemed as well established yet.
One problem is what is the major reason to discriminate L and T types.
Probably, T type may be defined by the appearance of the methane bands,
but it is now known that the CH$_{4}$ 3.3 $\mu$m bands appear at L5 
\citep{nol97}, and the 1.6 and/or 2.2$\mu$m bands appear at late L dwarfs 
\citep{nak01, mcl01}. Another problem is that the different L types 
show the same  infrared color because of the red limit of the infrared
colors at about L5-6. One possibility to resolve these
inconveniences may be to terminate the L type at about the present L5-6
and to allocate T type to the later objects. This is in accord with the
proposition by \citet{geb02} to define T type by the appearance
of the CH$_{4}$ 1.6 $\mu$m bands, and infrared colors can be unique for
L types as well as for T types. This is, however, not proposing to
change anything but for naming.

Although the L - T classification may represent a temperature sequence, 
accurate calibration of $ T_{\rm eff}$ against the spectral type is
not known yet. Our preliminary attempt (Sect.7.1) results in  the Table 2, 
from which L dwarfs may start at about $ T_{\rm eff}\approx 2000$\,K and may 
terminate at about  $ T_{\rm eff}\approx 1450$\,K. Then, T dwarf sequence 
may start at about  $ T_{\rm eff}\approx 1400$\,K and extends to  
$ T_{\rm eff}$ below 1000\,K. We must remember, however, the limitation 
of the color method to estimate $ T_{\rm eff}$ and an efficient method 
to relax this difficulty may be the
infrared flux method \citep{bla80} which utilize the bolometric
flux and an infrared flux that can be chosen to be relatively free from
the severe line blanketing effect. Unfortunately, however, it is difficult
to find such an infrared flux relatively free from the line blanketing 
effect in L and T dwarfs, even though it was still marginally possible
in M dwarfs \citep{tsu96a}. The most reliable method to estimate
$ T_{\rm eff}$ may be to measure bolometric flux of the ultracool dwarf
with known parallax as noted by \citet{bur02}. If $ T_{\rm eff}$ 
can be determined on purely empirical basis, we hope that our result 
in Table 2 can be used for testing the models.

\subsection{Observed Effects of Dust}

It appears to be no doubt that the spectra and colors of late M dwarfs 
and L dwarfs suffer a considerable effect of dust extinction. These results 
may have two implications: First, dust 
formed in the photosphere of cool dwarfs should be well mixed with gas
and uniformly cover the whole photosphere, since otherwise the effect of
dust on spectra and colours cannot be so large as observed. 
Second, the dust in the photosphere of cool dwarfs will survive for a long 
time, as long as the Hubble time, since otherwise the dust could  not be 
observed in so many cool dwarfs including the most late M dwarfs and L 
dwarfs. Our hypothesis that the dust grains formed in the photospheres
of  ultracool dwarfs should be very small ones which are in detailed 
balance with the gas will explain these results quite easily, since 
formation and destruction of such small grains will be repeating forever
everywhere in the pohotosphere near the dust condensation temperature.
On the other hand, it should be more difficult to consider a mechanism
to sustain large dust grains for a long time in the static photosphere.

In a more detailed  analysis of the spectra of dusty dwarfs, however,
a difficult problem appears, that is an additional parameter which 
we referred to as the critical temerature had to be introduced. 
This parameter is essentially related to the thickness of the dust 
cloud or, more correctly, to the dust column density of the cloud
in the observable photosphere. Unfortunately, direct determination of 
the dust column density in the observable photosphere is difficult, 
since dust shows no direct spectroscopic feature in the observed 
spectra. This is inherent difficulty in quantitative analyses of the 
spectra contaminated by dust and offers a formidarable problem.  
A more dificult case is the late T dwarfs in which the dust cloud 
may be situated too deep to give noticeable observable effect.
The presence of the warm dust deep in the photosphere cannot be known 
directly by observations of the cool T dwarfs themselves, but can only 
be inferred from the results on the warmer objects in which the dust 
cloud appears in the optically thin regime. For this reason, it may 
not be possible to prove nor to disprove the presence of the warm 
dust deep in the photosphere by the analyses of the spectra of the 
coolest T dwarf such as SDSS\,1624+00 \citep{nak00, lie00}.

So far, we showed that $T_{\rm cr}$ can be inferred from the red limits 
of the infrared colors by assuming that $T_{\rm cr}$ may be the same 
at least for the reddest late  L dwarfs (Sect.7.1). Although this 
assumption should most probably be too simplified, it may be difficult 
to determine $T_{\rm cr}$ for individual object. This is because colors 
(and any other observables) depend not only on $T_{\rm cr}$ but also 
on $T_{\rm eff}$, and  the effect of these two parameters could best 
be separated from the systematics on a large sample. Since 
$ T_{\rm cond}$ is higher for the models of lower $T_{\rm eff}$ (Fig.3), 
$ T_{\rm cr}$ may also be higher for the cooler dwarfs. We have prepared
the model photospheres that may cover the possible range of $T_{\rm cr}$
and we hope that they can be used for estimating $T_{\rm cr}$ in
ultracool dwarfs of different $T_{\rm eff}$'s.
For example, dust may be formed in late M dwarfs, but the dust cloud may be
too thin for $T_{\rm cr} \approx 1800$\,K, since $T_{\rm cond}$ may
already be close to 1800\,K in late M dwarfs (Fig.3). Clearly, we cannot apply 
the same value of $T_{\rm cr}$ to M dwarfs as for L dwarfs, and we hope that
some methods to estimate  $T_{\rm cr}$ in M dwarfs can be developed.  

Another problem related to dust is  if it may induce some meteorological 
phenomena which may explain the unknown type of observed light variations 
in some ultracool dwarfs \citep[e.g.][]{bai01, mart01}. Even if the dust 
cloud composed of the small grains is quite stable as noted above, it 
may be possible that some inhomogeneities develope in the cloud.
Also, we do not yet consider the fate of the large grains which we
assumed to have precipitated below the observable photosphere and
might have evaporated there. It may be possible that the large grains
do not necessarily precipitate so easily but remain in the photosphere
as real ``clouds'' which are more familiar to us on the Earth.
Such ``clouds'' should have only a small filling factor of the
stellar surface, since otherwise their effect should be observable
in T dwarfs as well. We have not yet considered the fate of the large 
grains that have segregated from the gaseous mixture and  we still have 
many unsolved problems  about dust in ultracool dwarfs.

\section{CONCLUDING REMARKS}

In general, the purpose of modeling is to understand  complicated 
astronomical phenomena by simple physical principles with least ad-hoc
assumptions. In this respect, the model stellar photosphere has been one 
of the successful cases within the framework of  the so-called classical 
model stellar photosphere. In fact, extensive grids of the classical 
model photospheres have already been available for most stars including 
O through M types  and they showed reasonable successes in the 
interpretations and analyses of observed data. 
We simply extended the method of the model stellar photosphere
to effective temperatures as low as 800\,K by considering the effect of 
dust. This could be realized by an additional assumption that
only small dust grains (smaller than about 0.01 $\mu$m)  
can be sustained in the photosphere, resulting in a dust cloud deep in the 
photosphere.  We should like to emphasize that
this assumption can be regarded as being well supported by the fact that
major observations on ultracool dwarfs could be understood reasonably 
well by our cloudy models based on this assumption. 

This fact in turn provides an answer to the problem on how dust 
forms in the photospheric environment; dust forms and survives rather 
deep in the photosphere near the dust condensation temperature and not 
in the cooler surface region. Thus, the major conclusion which we 
learned from the extensive observations on L and T dwarfs is that the 
dust cloud should be formed deep in the photospheres of these ultracool 
dwarfs and this result may be applied to modeling the photosphere with 
dust in general. We applied this result to the modeling of the 
photosphrere of substellar brown dwarfs with $T_{\rm eff} \ga 800$\,K 
in this paper, and showed that the photospheric structure of substellar 
objects can be modeled by applying the methodology of stellar modeling
without incorporating the complexities of the planetary model atmospheres.
Probably, our cloudy model may be the most simple model that
incorporated the cloud formation, and hopefully provide a clue
for unified modeling of the cloudy  atmospheres of stars, 
substellar objects, and giant planets.

\acknowledgements
I thank Tadashi Nakajima for helpful discussion throughout this work
and for reading the manuscript with valuable comments.
My thanks are also due to Aleksandra Borysow for making available
her codes to compute the absorption coefficients of H$_{2}$ CIA and
for helpful correspondences. Finally, I am indebted to the referee, 
Katharina Lodders, for critical reading of the text and for helpful 
suggestions on the thermochemistry. This work was  supported by the 
grant-in-aid No.11640227 of JSPS, and carried out with the use of
the facilities at the Astronomical Data Analysis Center of
NAO, which is an inter-university research institute of astronomy 
operated by Ministry of Education \& Science.

\appendix

\section{MOLECULAR OPACITY DATA} 

\subsection{ Infrared Bands of H$_{2}$O}

Until recently, we have been using the experimental data for the 
straight mean absorption cross-sections and the mean line separation 
by Ludwig (1967). In the present work, the mean absorption cross-sections 
over 25 cm$^{-1}$ interval are generated from the HITEMP database 
\citep{rot97}, which is a high temperature adjunct 
to the well  known HITRAN database for atmospheric lines \citep{rot98}.
As to the mean line separation, we have also been using Ludwig's data, but 
we found that the resulting spectra based on the
band model method using the mean line separation derived from the
fine structure parameter by \citet{lud67} show rather poor 
agreement with those by the detailed line-by-line method using the
HITEMP database. Accordingly, we introduced correction factors to the 
Ludwig's data so that the spectra based on the band model agree with 
those by the detailed line-by-by method.

\subsection{ Infrared Bands of Polyatomic Molecules}
 
Although spectroscopic data at low temperatures are relavively well known 
for some  polyatomic molecules thanks to the recent spectroscopic 
databases such as  HITRAN \citep{rot98} and GEISA \citep{jac99}, little 
is known on pectroscopic properties at elevated temperatures except for 
H$_{2}$O. Thus, only practical approach may be to apply 
the band model in its most simple form - JOLA -  based on the 
rigid-rotator harmonic osscillator model \citep{tsu94}.
Even the application of this simple method is not necessarily
easy because of the lack of the intensity data for some bands. 

CH$_{4}$: 
This spherical top molecule has 4 fundamentals,
of which $\nu_3$ and $\nu_4$  are infrared active (but $\nu_1$ and $\nu_2$ 
are also weakly active). We have considered 14 band systems for which 
spectroscopic and intensity data are summarized in Table 3. 
Unfortunately, some bands are still missing in the region shortward of
1.6 $\mu$m. In Fig.14, the resulting absorption coefficient of methane 
at $T = 296$\,K based on the band model is compared with that based on the 
GEISA database. The test is limited to the room temperature at which 
the GEISA database can be valid, but the band model opacity can in 
principle be applied to the higher temperatures without limit.  

NH$_{3}$ and PH$_{3}$:
These symmetric top molecules have 4 fundamentals for which intensity data
are available. But intensity data for overtones are definitely 
meager and we could include only one overtone for NH$_{3}$. 
Thus,  only five and four band systems  are included for NH$_{3}$ and 
PH$_{3}$, respectively, with the intensity data summarized in Table 3. 
In addition to these, the pure rotation transitions are
included with the known electric dipole moments \citep{nel67}.

H$_{2}$S:
This molecule is quite similar to H$_{2}$O, but the intensities of the 
infrared bands are relatively small compared with H$_{2}$O. We selected 
relatively strong 34 band systems from the intensity data evaluated 
by \citet{sen89}. The pure rotation transition is also included.

\subsection{ Infrared Bands of Diatomic Molecules}

The band model parameters - the mean absorption cross-section and
the mean line separation -  are evaluated from the spectroscopic data  
for the fundamental, first and second overtone bands of CO, OH, and SiO. 
The spectroscopic and intensity data are relatively well fixed except 
for the recent revision of oscillator strengths of SiO bands \citep{lan93}.
The band model opacity is tested by comparing the emergent fluxes based 
on the band model and those evaluated by the use of detailed 
linelist \citep{tsu94}.

\subsection{ Electronic Bands of Diatomic Molecules}

We evaluated the mean absorption cross-sections and mean line 
separations for the following 5 molecules. The adopted $f$-values are 
summarized in Table 4.

TiO: We considered 8 systems of TiO shown in Table 4. The
$f$--values for $\delta$ and $\phi$ systems are by \citet{dav86},
and those for other 6 systems are based on the
direct measurements of the radiative lifetimes by \citet{hed95}. 

VO: Some details on the Yellow-Green $(C-X)$ System are known \citep{har71}. 
The importance of VO infrared bands as opacity sources 
was first noted by Brett (1990), and we apply the $f$-values of $A-X$ and 
$B-X$ systems based on the recent lifetime measurements \citep{kar97}.

FeH: A detailed spectroscopic analysis of the $F-X$ system is done by 
\citet{phi87}. No experimental $f$-value is known and we applied the 
{\it ab initio} results by \citet{lan90}. We used the dissociation
energy of 1.63 eV by the mass spectrometer experiment \citep{sch91}.

CaH : Intensity and spectroscopic data are not known well, and we had to 
use  an astronomical $f$-value for $A-X$ system \citep{mou76}.

MgH : We apply the {\it ab initio f-}value for $A-X$ system \citep{kir79}.

\clearpage

\begin{table}
\caption{CHEMICAL COMPOSITION\tablenotemark{a}}
\begin{tabular}{ccccccc}
\noalign{\bigskip}
\tableline\tableline
\noalign{\bigskip}
Atomic No. & El. & {log $A_{\rm El}$} &  ~~~~~ &
Atomic No. & El. & {log $A_{\rm El}$} \\
\noalign{\bigskip}
\tableline
\noalign{\bigskip}
 1  &  H  & 12.00     &  & 20  &  Ca & 6.36 \\
 2  &  He  & 10.99   &  & 21  &  Sc & 3.10 \\
 3  &  Li  & 3.31     &  & 22  &  Ti  & 4.99 \\
 4  &  Be  & 1.15     &  & 23  &  V   & 4.00 \\
 5  &  B   & 2.60     &  & 24  &  Cr  & 5.67 \\
 6  &  C   & 8.60\tablenotemark{b}     &  & 25  &  Mn  & 5.39 \\
 7  &  N   & 8.00\tablenotemark{c}     &  & 26  &  Fe  & 7.51\tablenotemark{e} \\
 8  &  O   & 8.92\tablenotemark{d}    &   & 28  &  Ni  & 6.25 \\
 9  &  F   & 4.56    &   & 29  &  Cu  & 4.21 \\
 11 & Na   & 6.33   &    & 35  & Br   & 2.63 \\
 12 & Mg   & 7.58   &    & 37  & Rb  & 2.69  \\
 13 & Al   & 6.47   &    & 38  & Sr  & 2.90 \\
 14 & Si   & 7.55  &     & 39  & Y   & 2.24 \\
 15 &  P   & 5.45  &     & 40  & Zr  & 2.60 \\
 16 &  S   & 7.21  &     & 53  &  I  & 1.51 \\ 
 17 & Cl   & 5.50  &     & 56  &  Ba & 2.13 \\
 19 & K   &  5.12  &     & 57  & La  & 1.22 \\
\noalign{\bigskip}
\tableline
\end{tabular}

\tablenotetext{a}{based on Anders \& Grevesse (1989) except for those 
footnoted}
\tablenotetext{b}{Grevesse et al.(1991)}
\tablenotetext{c}{Grevesse et al.(1990)}
\tablenotetext{d}{recent value of 8.69 by Allende Prieto et al.(2001) is
also tried in some models}
\tablenotetext{e}{Bi\'emont et al.(1991); Holweger et al.(1991)}

\end{table}

\begin{table}
\caption{COLOR INDECES AND EFFECTIVE TEMPERATURES}
\begin{tabular}{lclclcll}
\noalign{\bigskip}
\tableline\tableline
\noalign{\bigskip}
Sp. type & ${J-K}$ & ~${T_{\rm eff}}$~~~~~ &
  ${J-H}$ & ~${T_{\rm eff}}$~~~~~ &
 ${H-K}$ & ~${T_{\rm eff}}$~~~~~ & ${T_{\rm eff}}$(mean)\\
\noalign{\bigskip}
\tableline
\noalign{\bigskip}
L0  &  1.0 & 1970\,K & 0.5 & 2150\,K & 0.45 &  1960\,K & 2030\,K \\
L3  &  1.5 & 1780 & 0.85 & 1770 & 0.7 &  1740 & 1760 \\
L6  & 1.8 & 1520 & 1.0 & 1510 & 0.8 & 1540 & 1520 \\
L9  &  1.5  & 1450 & 0.9 & 1450 & 0.55 & 1440 & 1450 \\
T2  &  0.8  & 1320 & 0.5 & 1280 & 0.25 & 1340 & 1310 \\
T5  &  -0.05 & 1050 & -0.05 & 1110 & 0.0 & ~800 & ~990 \\  
\noalign{\bigskip}
\tableline
\end{tabular}
\end{table}

\clearpage

\begin{table}
\caption{ INTENSITY DATA OF POLYATOMIC MOLCULES}
\begin{flushleft}
\begin{tabular}{lcccc}
\tableline\tableline
\noalign{\bigskip}
Molecule & ${ n_{1}n_{2}n_{3}n_{4}}$ & $\omega$(cm$^{-1}$) & 
 $f_{n_{1}n_{2}n_{3}n_{4}}^{0}$  & Ref.     \\
\noalign{\bigskip}
\tableline
\noalign{\bigskip}
   NH$_{3}$   &  0 1 0 0 &  950  &  $ 2.39 \times 10^{-5}$  &  1  \\
              &  0 0 0 1 & 1630  &  $ 4.65 \times 10^{-6}$  &  1  \\
              &  0 2 0 0 & 1739  &  $ 4.72 \times 10^{-8}$  &  2  \\
              &  1 0 0 0 & 3337  &  $ 1.03 \times 10^{-6}$  &  3  \\
              &  0 0 1 0 & 3448  &  $ 6.04 \times 10^{-7}$  &  3  \\
\noalign{\bigskip}
   PH$_{3}$   &  0 1 0 0 &  992  &  $ 3.45 \times 10^{-6}$  &  4  \\
              &  0 0 0 1 & 1122  &  $ 4.29 \times 10^{-6}$  &  4  \\
              &  1 0 0 0 & 2323  &  $ 1.09 \times 10^{-5}$  &  4  \\
              &  0 0 1 0 & 2328  &  $ 1.09 \times 10^{-5}$  &  4  \\
\noalign{\bigskip}
   CH$_{4}$   &  0 0 0 1 & 1311  &  $ 5.65 \times 10^{-6}$  &  5  \\
            &  0 1 0 0 & 1533  &  $ 1.01 \times 10^{-7}$  &  6  \\
             &  0 0 0 2 & 2622  &  $ 1.13 \times 10^{-7}$  &  6  \\
             &  0 1 0 1 & 2844  &  $ 9.05 \times 10^{-7}$  &  6  \\
             &  1 0 0 0 & 2916  &  $ 1.31 \times 10^{-9}$  &  7  \\
\noalign{\smallskip}
             &  0 0 1 0 & 3019  &  $ 1.21 \times 10^{-5}$  &  8  \\
             &  0 2 0 0 & 3067  &  $ 3.85 \times 10^{-8}$  &  7  \\
             &  0 0 0 3 & 3932  &  $ 3.43 \times 10^{-8}$  &  9  \\
             &  0 1 0 2 & 4155  &  $ 3.43 \times 10^{-8}$  &  9  \\
             &  1 0 0 1 & 4227  &  $ 3.16 \times 10^{-7}$  &  10  \\
\noalign{\smallskip}
             &  0 0 1 1 & 4330  &  $ 4.61 \times 10^{-7}$  &  7  \\
             &  0 1 1 0 & 4553  &  $ 7.07 \times 10^{-8}$  &  7  \\
             &  0 0 1 2 & 5641  &  $ 1.39 \times 10^{-9}$  &  7  \\
             &  2 0 0 0 & 5832  &  $ 7.49 \times 10^{-8}$  &  11  \\
             &  1 0 1 0 & 5935  &  $ 7.49 \times 10^{-8}$  &  11 \\
\noalign{\smallskip}
             &  0 0 2 0 & 6039  &  $ 7.49 \times 10^{-8}$  &  12  \\
\noalign{\bigskip}
\tableline
\noalign{\bigskip}
\end{tabular}
\tablerefs{
 (1) Kim 1985;
 (2) Brown et al. 1987;
 (3) Pine \& Dang-Nhu 1993;
 (4) McKean \& Schatz 1956;
 (5) Restelli \& Cappellani 1982;
 (6) MaClatchey et al. 1973;
 (7) Rothman et al. 1998;
 (8) Dang-Nhu et al. 1979;
 (9) Brown 1988;
 (10) Brown 1982;
 (11) $f$ value is assumed to be the same as that of 2$\nu_{3}$ band;
 (12) Margolis 1988.
}
\end{flushleft}
\end{table}

\clearpage

\begin{table}
\caption{THE $f-$VALUES OF THE ELECTRONIC BANDS}
\begin{flushleft}
\begin{tabular}{llrlc}
\tableline\tableline
\noalign{\bigskip}
Molecule & Transition & $\omega_{00}$~~~ &   ~$f_{00} $ & Ref. \\
         &           &   (cm$^{-1})$  &            &    \\
\noalign{\bigskip}
\tableline
\noalign{\bigskip}
TiO   & $A^{3}\Phi$-$X^{3}\Delta (\gamma)$ & 14,163 & 0.112  &  1  \\
     & $B^{3}\Pi$-$X^{3}\Delta (\gamma^{'})$ & 16,219 & 0.100  &    1  \\
      & $C^{3}\Delta$-$X^{3}\Delta (\alpha)$ & 19,424 & 0.073 &    1  \\
      & $E^{3}\Pi$-$X^{3}\Delta (\epsilon)$ & 11,885 & 0.012    &    1  \\
      & $b^{1}\Pi$-$a^{1}\Delta (\delta)$ & 11,319 & 0.039      &    2  \\
      & $c^{1}\Phi$-$a^{1}\Delta (\beta)$ & 17,890 & 0.26       &    1  \\
      & $f^{1}\Delta$-$a^{1}\Delta$ & 19,140 & 0.061             &    1  \\
      & $b^{1}\Pi$-$d^{1}\Sigma (\phi)$ & 9,100 & 0.03        &    2  \\
\noalign{\bigskip}
VO   & $A^{4}\Pi$-$X^{4}\Sigma^{-}$  & 9,494 & 0.009     &    3  \\
     & $B^{4}\Pi$-$X^{4}\Sigma^{-}$ & 12,606 & 0.16        &    3 \\
     & $C^{4}\Sigma^{-}$-$X^{4}\Sigma^{-}$ & 17,420 & 0.009     &    4  \\
\noalign{\bigskip}
FeH  & $F^{4}\Delta$-$X^{4}\Delta$ & 9,995 & 0.013 &    5  \\
\noalign{\bigskip}
CaH  & $A^{2}\Pi$-$X^{2}\Sigma$ & 14,394  &  0.059       &    6  \\
     & $B^{2}\Pi$-$X^{2}\Sigma$ & 15,754  &  0.113        &    7  \\
\noalign{\bigskip}
MgH  & $A^{2}\Pi$-$X^{2}\Sigma^{+}$ & 19,288  & 0.161       &   8  \\
\noalign{\bigskip}
\tableline
\end{tabular}
\end{flushleft}
\tablerefs{
 (1) Hedgecock et al. 1995;
 (2) Davis et al. 1986;
 (3) Karlsson 1997;
 (4) Harrington et al. 1970;
 (5) Langhoff \& Bauschlicher 1990;
 (6) Mould 1976;
 (7) Klynning 1982;
 (8) Kirby et al. 1979.}
\end{table}

\clearpage

\begin{figure}
\epsscale{0.75}
\plotone{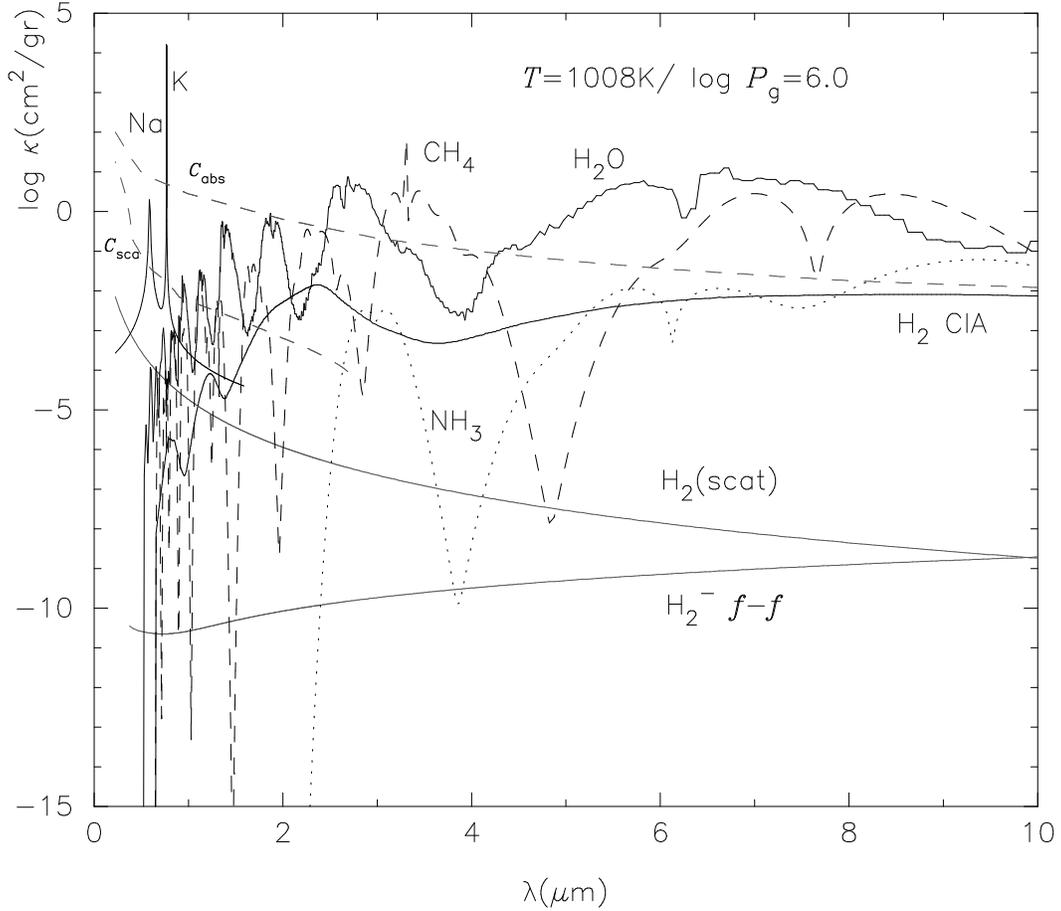}
\caption {
 The extinction coefficients (per gram of stellar material)
of major opacity sources in the solar composition mixture at $T=1,008$K
and log $P_{g}$=6.0. The absorption and scattering coefficients of
iron particles ($r_{\rm gr} = 0.01 \mu$m) are also shown by $C_{\rm abs}$ and
$C_{\rm sca}$, respectively, although dust grains may have segregated
from the gaseous mixture and may not be effective
as sources of opacity at the physical condition shown here.
}
\label{Fig1}
\end{figure}

\clearpage

\begin{figure}
\epsscale{1.0}
\plottwo{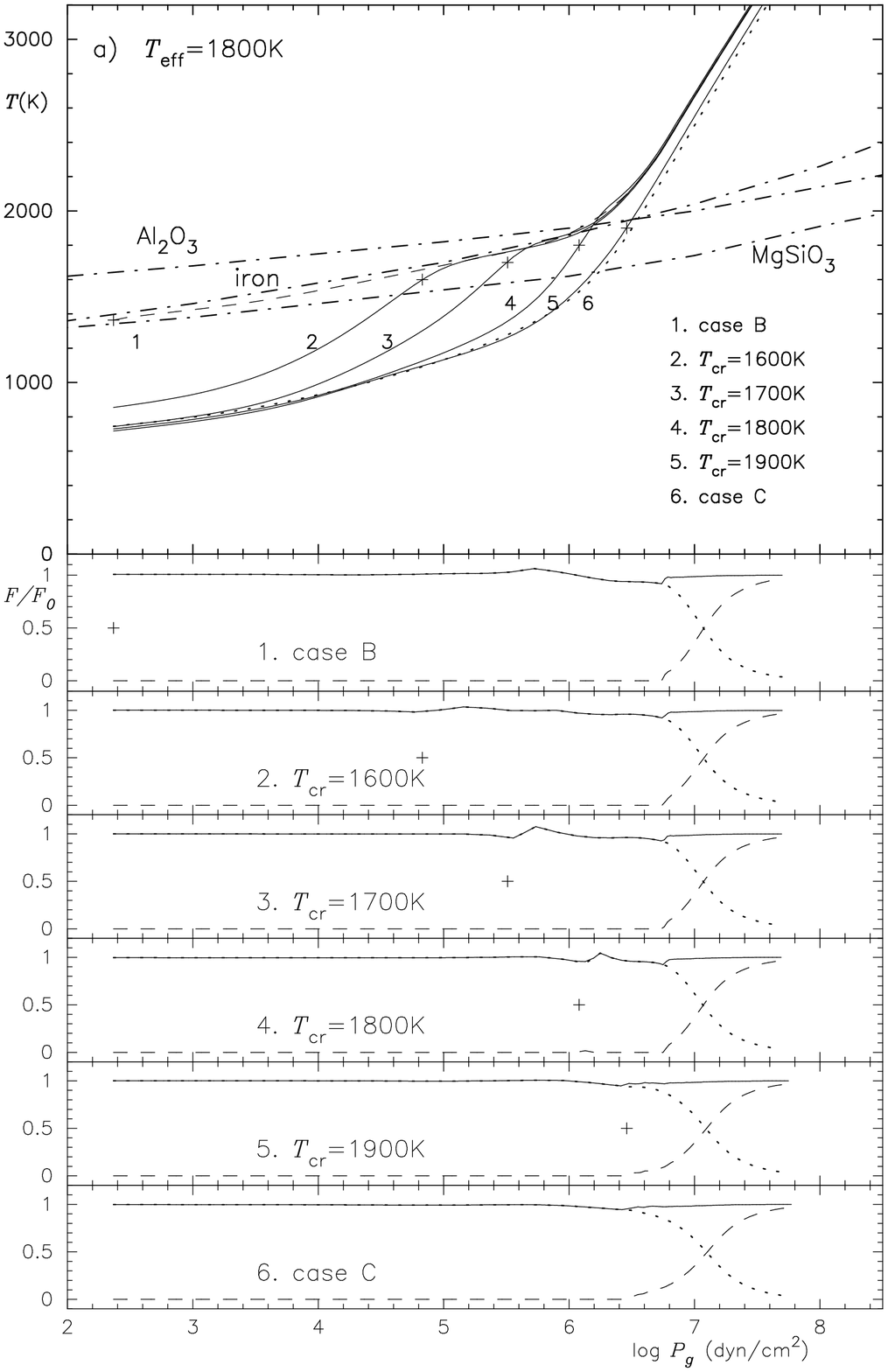}{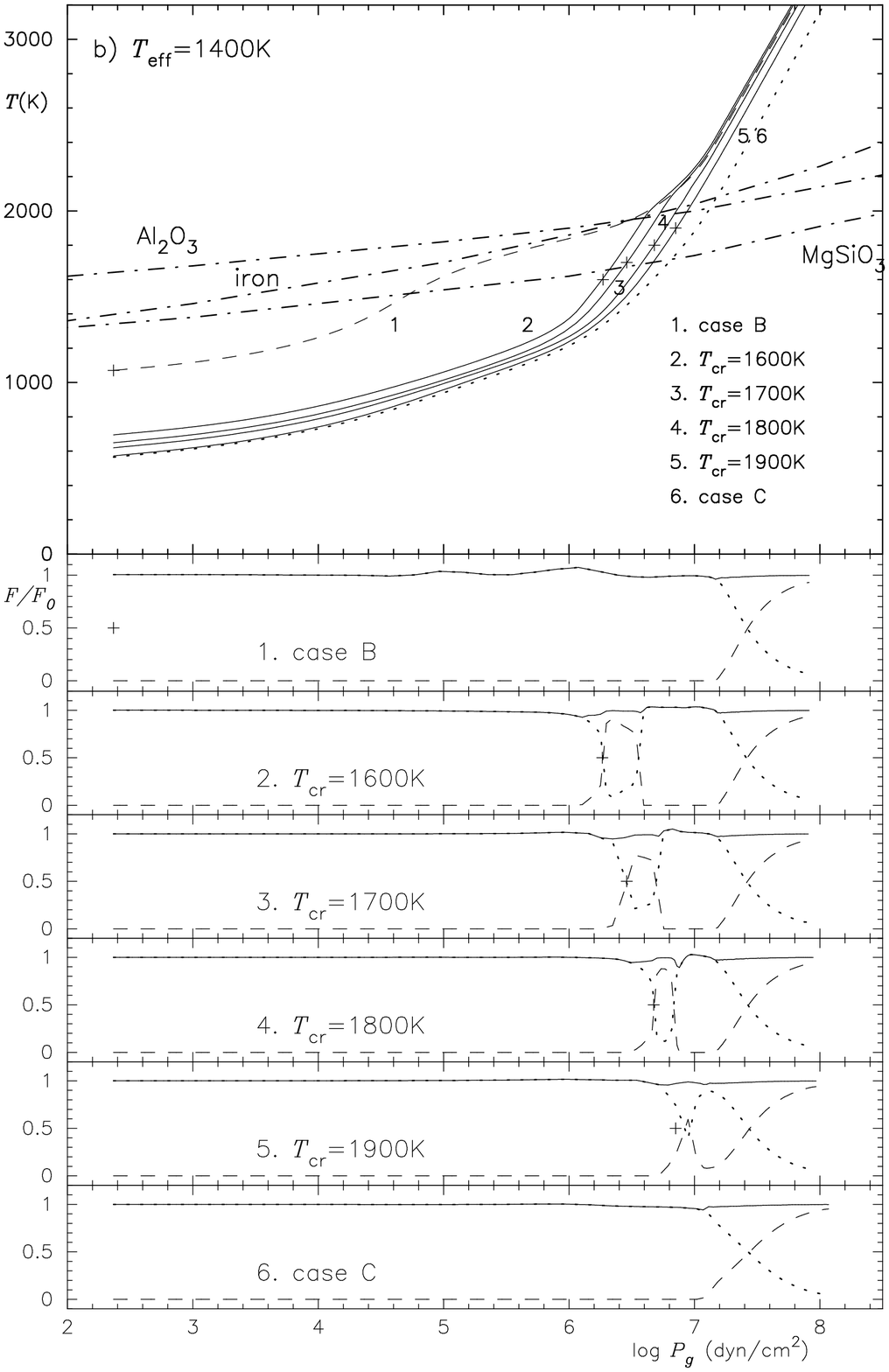}
\caption {
a) Non-grey model photospheres of $T_{\rm eff}$ = 1800K (log $g$ =5.0, 
$v_{\rm micro}=$1 km s$^{-1}$ and the solar metallicity) are shown in the 
upper panel for six values of the critical temperatures; $T_{\rm cr} 
= T_{0}$ (case B), 1600, 1700, 1800, 1900, and $T_{\rm cond}$ (case C). 
The solid lines illustrate the cloudy  models with $T_{\rm cr}$'s indicated
and the plus signs denote the upper boundary of the cloud (where $T = 
T_{\rm cr}$). The dashed and dotted lines show the extreme limiting 
cases B and C, respectively. The dot-dashed curves are the dust 
condensation lines for corundum, iron, and enstatite. The lower six 
panels show the  radiative, convective, and total fluxes normalized by 
$\sigma T_{\rm eff}^4/\pi$ by the dotted, dashed, and solid lines, 
respectively, for six values of the critical temperatures. 
b) The same as a) but for the case of $T_{\rm eff}$ = 1400K. Note a
drastic change of the convective structure compared with the case of 
$T_{\rm eff}$ = 1800K.
}
\label{Fig2}
\end{figure}

\clearpage

\begin{figure}
\epsscale{0.75}
\plotone{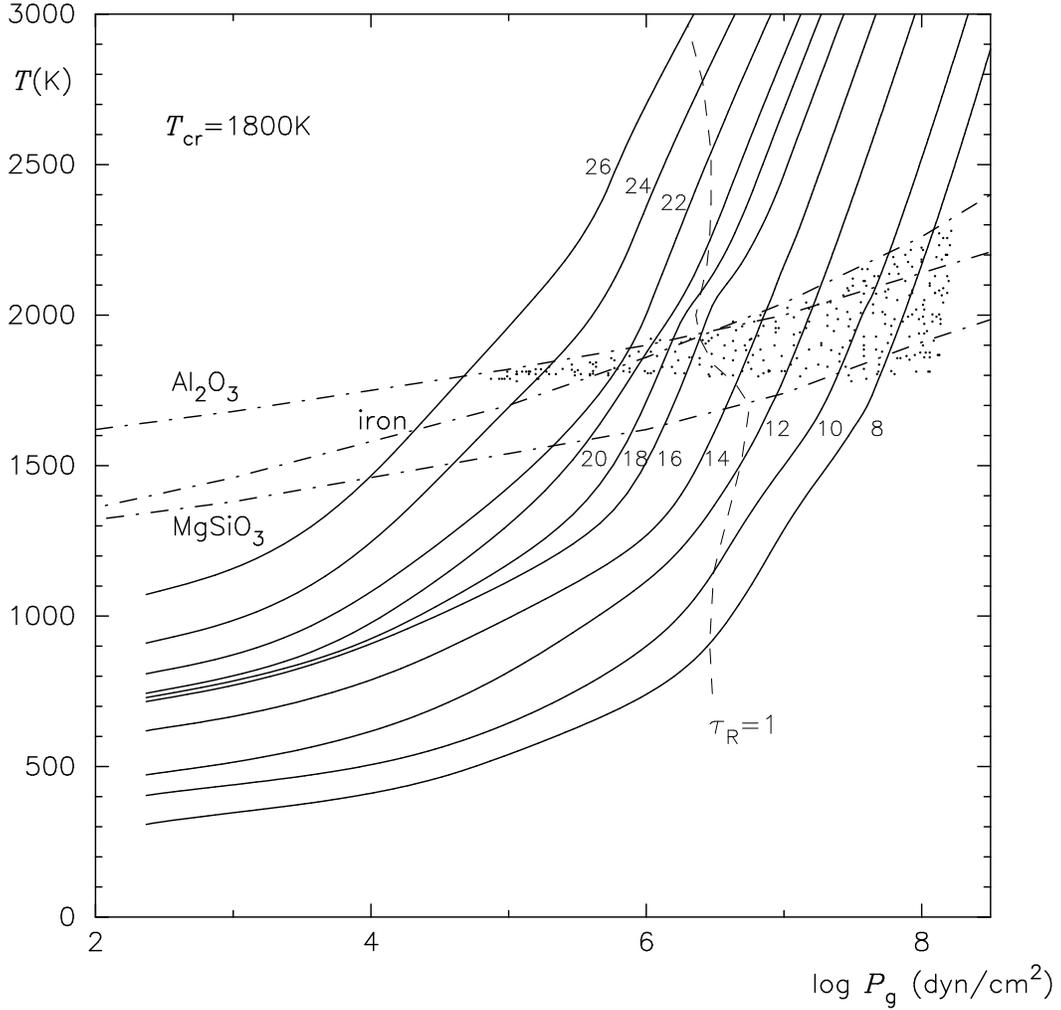}
\caption {
The cloudy model photospheres with $T_{\rm cr} = 1800$\,K for $T_{\rm eff}$ 
from  = 800\,K to 2600\,K  by  a step of $\Delta T_{\rm eff} = 200$K 
(log $g$ =5.0, $v_{\rm micro}=$1 km s$^{-1}$ and the solar metallicity).
The numbers attached are $T_{\rm eff}$ in units of 100 Kelvin and the 
cloud zone is shown by the dotted area. The locus where the Rosseland mean 
optical depth unity is shown by the dashed line. 
Some details of the radiative/convective structures  are shown in Fig.5.  
}
\label{Fig3}
\end{figure}

\clearpage

\begin{figure}
\epsscale{0.75}
\plotone{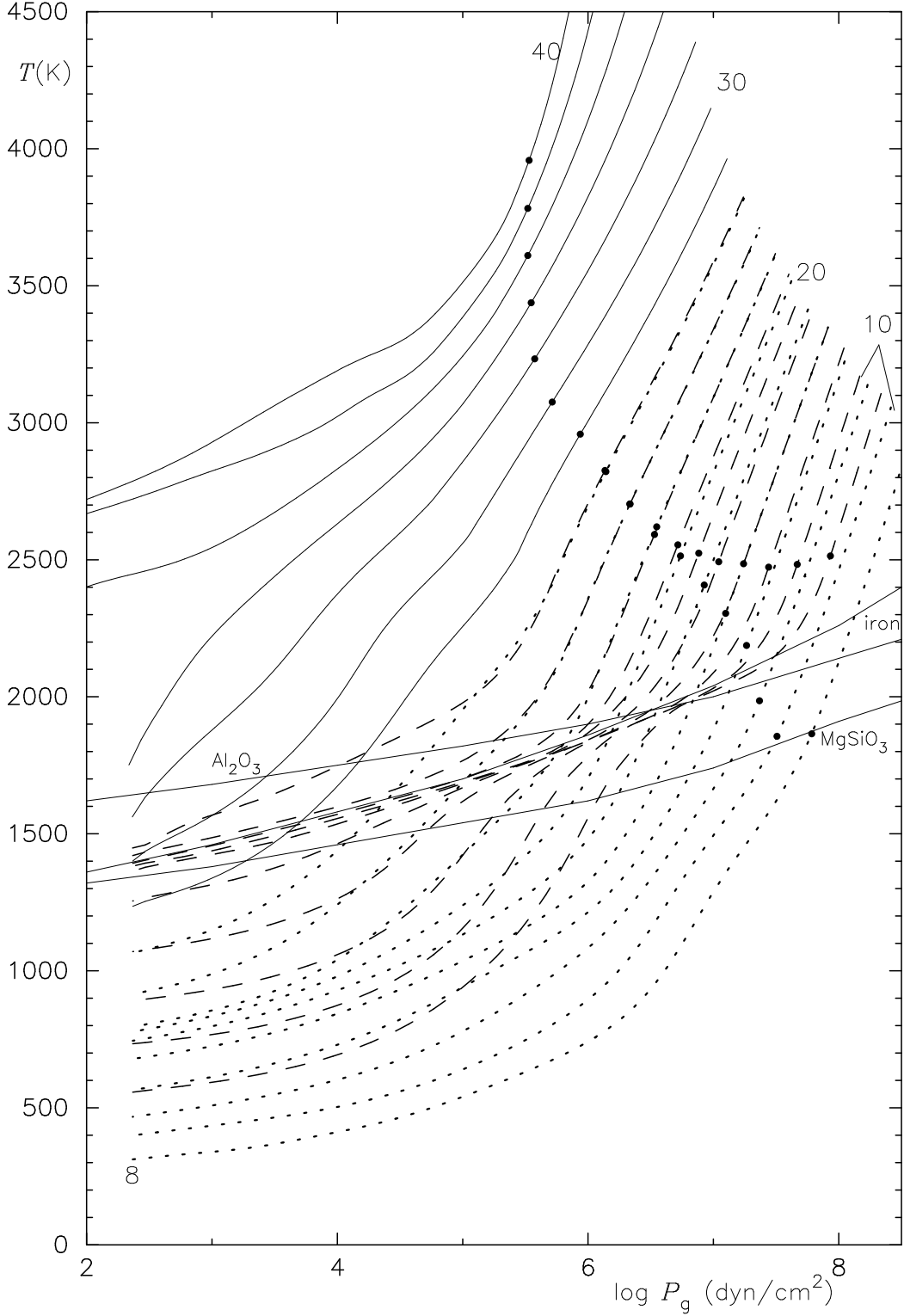}
\caption {
The extreme limiting cases of dusty (case B) and dust segregated 
(case C) models are shown by the dashed and dotted lines, respectively, 
for  $T_{\rm eff}$ from 800\,K to 2600\,K  by a step of $\Delta T_{\rm eff} 
= 200$K. The case C  models are extended to the dust-free regime with
$T_{\rm eff}$ between 2800\,K and  4000\,K and  shown by
the solid lines. The filled circles are where  $F_{\rm rad} = F_{\rm cnv}$ 
and denote the transition from the radiative to convective zones.
Other details are the same as the Fig.3 legend.
}
\label{Fig4}
\end{figure}

\clearpage

\begin{figure}
\epsscale{1.0}
\plottwo{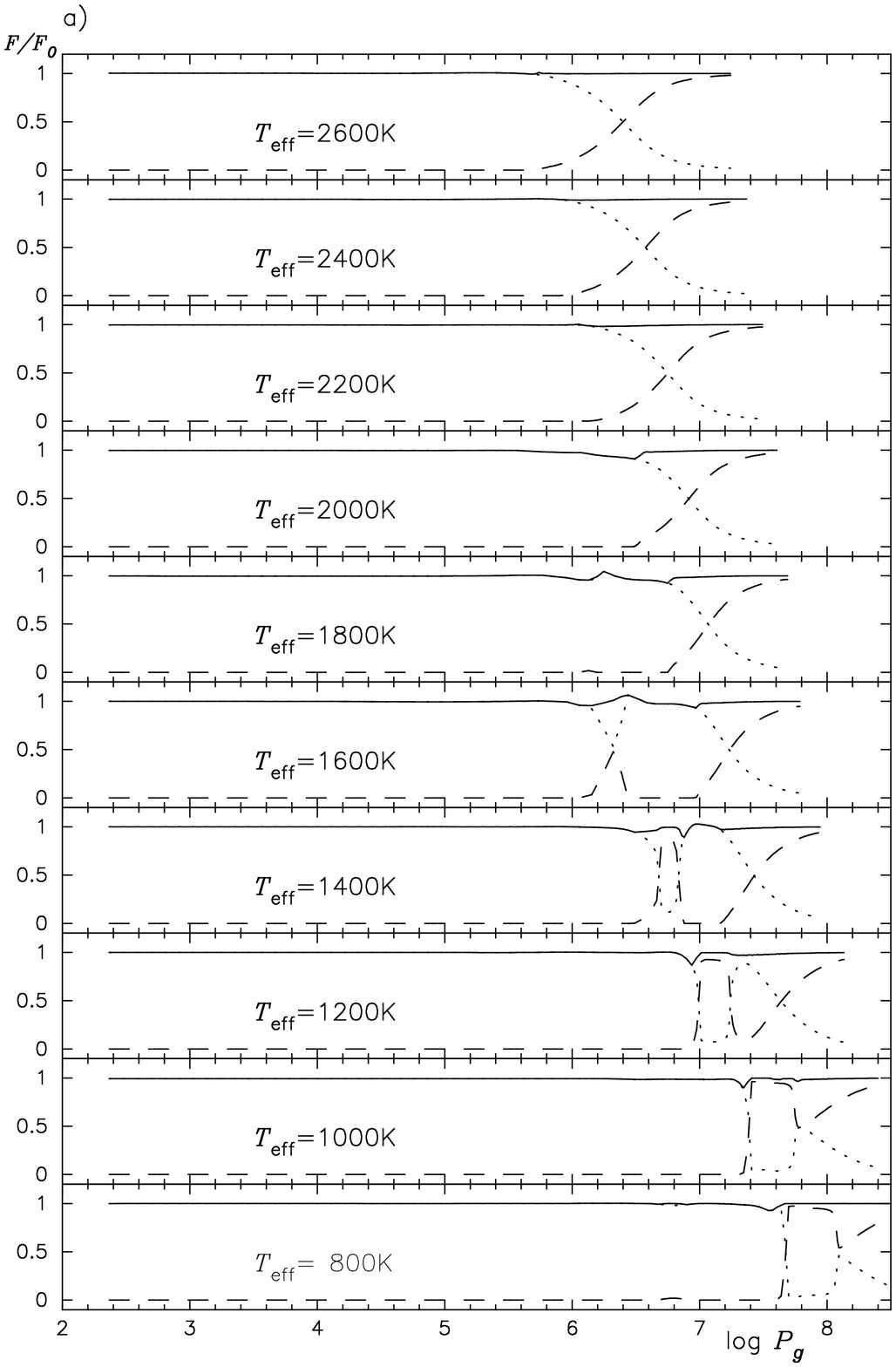}{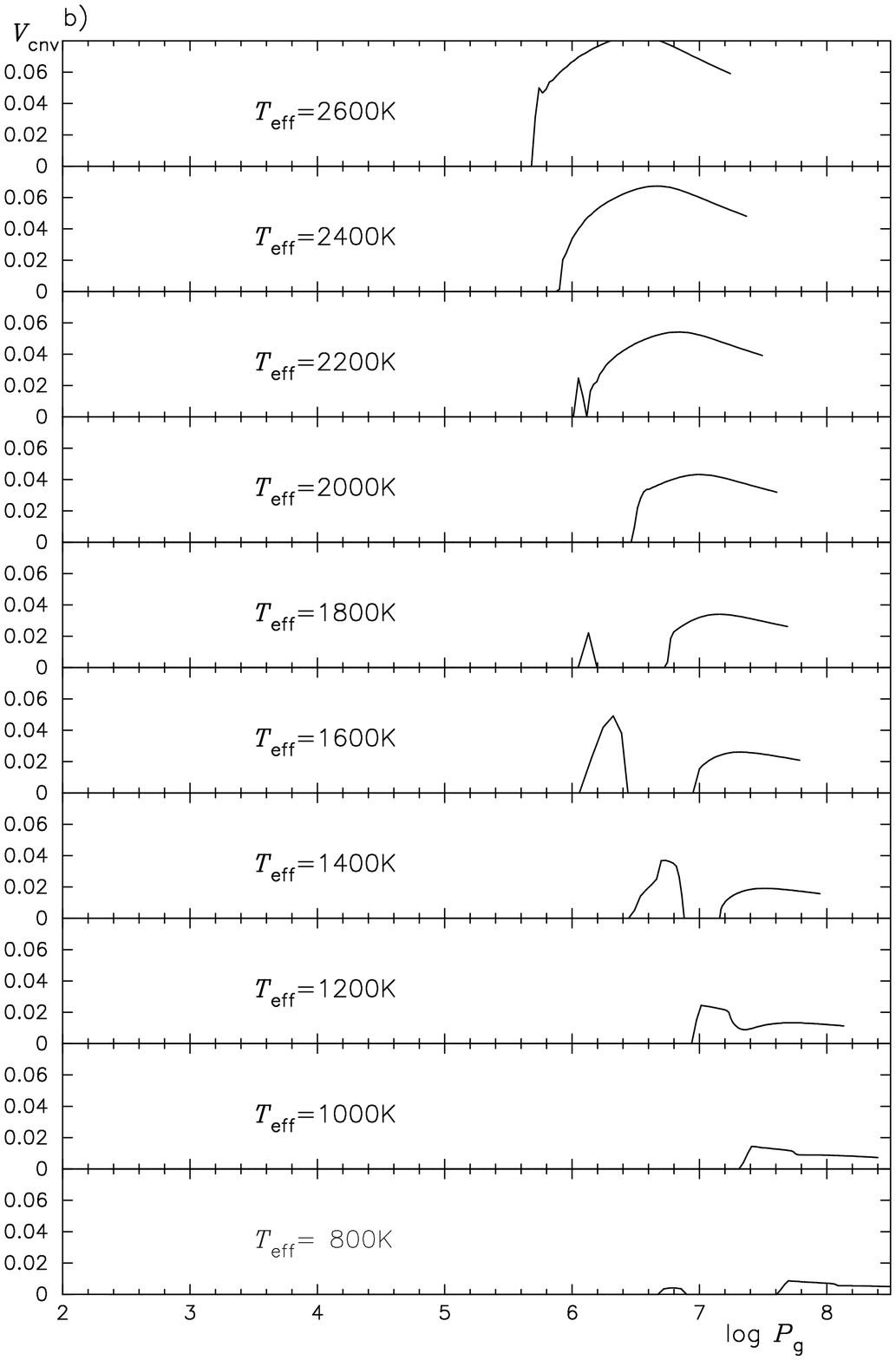}
\caption {
a) The  radiative, convective, and total fluxes normalized by 
$\sigma T_{\rm eff}^4/\pi$ are shown by the dotted, dashed, and solid lines, 
respectively, for  the cloudy model photospheres 
with $T_{\rm cr} = 1800$\,K for  $T_{\rm eff}$ from 800\,K to 2600\,K.
b) The convective velocity in units of km s$^{-1}$ for the same models
shown in a). 
}
\label{Fig5}
\end{figure}

\clearpage

\begin{figure}
\epsscale{0.75}
\plotone{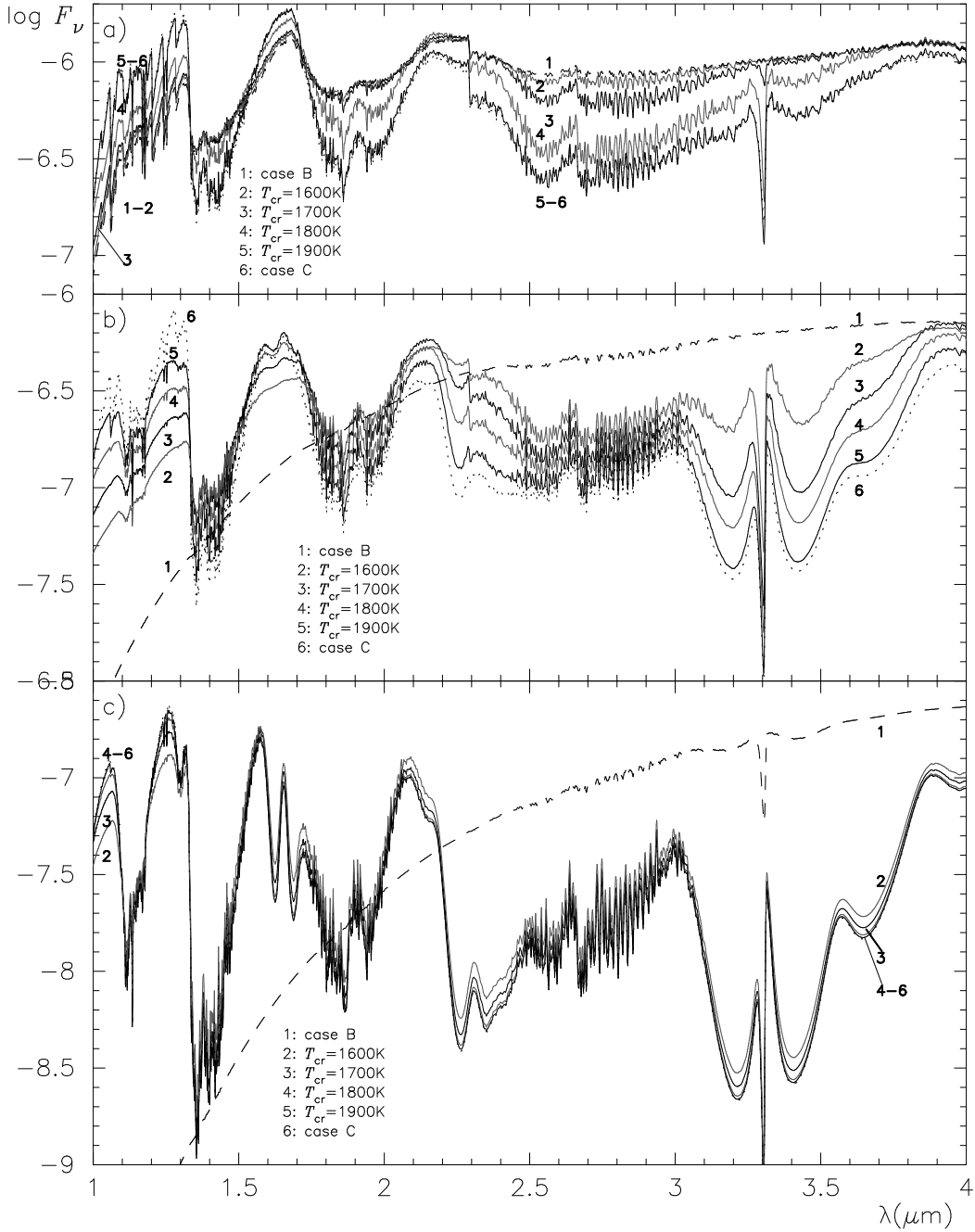}
\caption {
a) Predicted spectra from the models of the same $T_{\rm eff}$ 
(log $g$ =5.0, $v_{\rm micro}=$1 km s$^{-1}$ and the solar metallicity) 
for six values of the critical temperatures; $T_{\rm cr} = T_{0}$ (case B),
1600, 1700, 1800, 1900, and $T_{\rm cond}$ (case C). 
The  dashed, solid (black and grey), and dotted lines illustrate   
model B,  cloudy models, and model C, respectively.
a) $T_{\rm eff}$ = 1800K.
b) $T_{\rm eff}$ = 1400K. 
c) $T_{\rm eff}$ = 1000K. 
}
 \label{Fig6}
\end{figure}

\clearpage

\begin{figure}
\epsscale{0.75}
\plotone{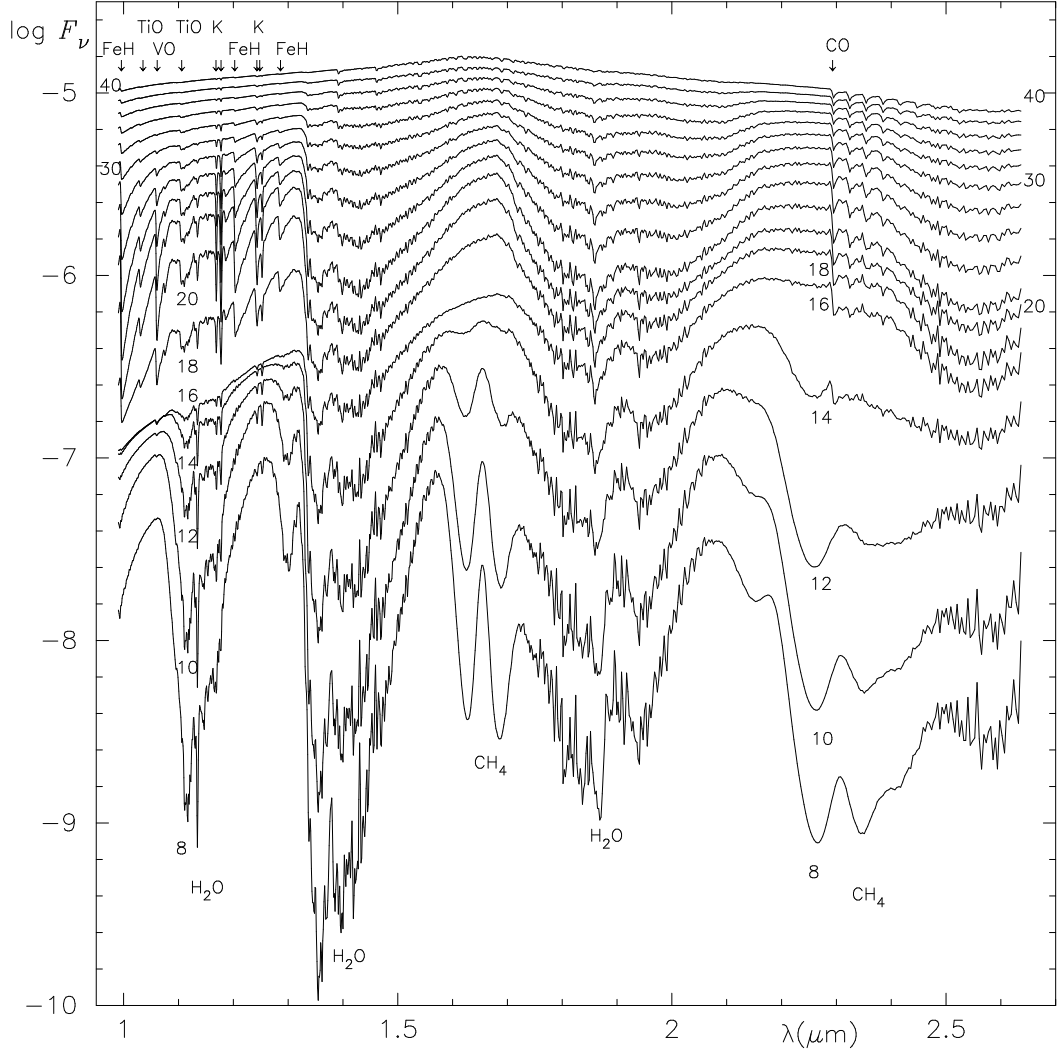}
\caption {
The spectra between 1.0 and 2.6 $\mu$m predicted from the cloudy models 
with $T_{\rm cr} = 1800$\,K for $T_{\rm eff}$  between 800 
and 2600\,K and those from the dust-free models for $T_{\rm eff}$  between 
2800 and 4000\,K. The step of $ T_{\rm eff}$  is 200\,K, and 
$T_{\rm eff}$ in  units of 100\,Kelvin are indicated on some spectra. 
}
\label{Fig7}
\end{figure}

\clearpage

\begin{figure}
\epsscale{1.0}
\plottwo{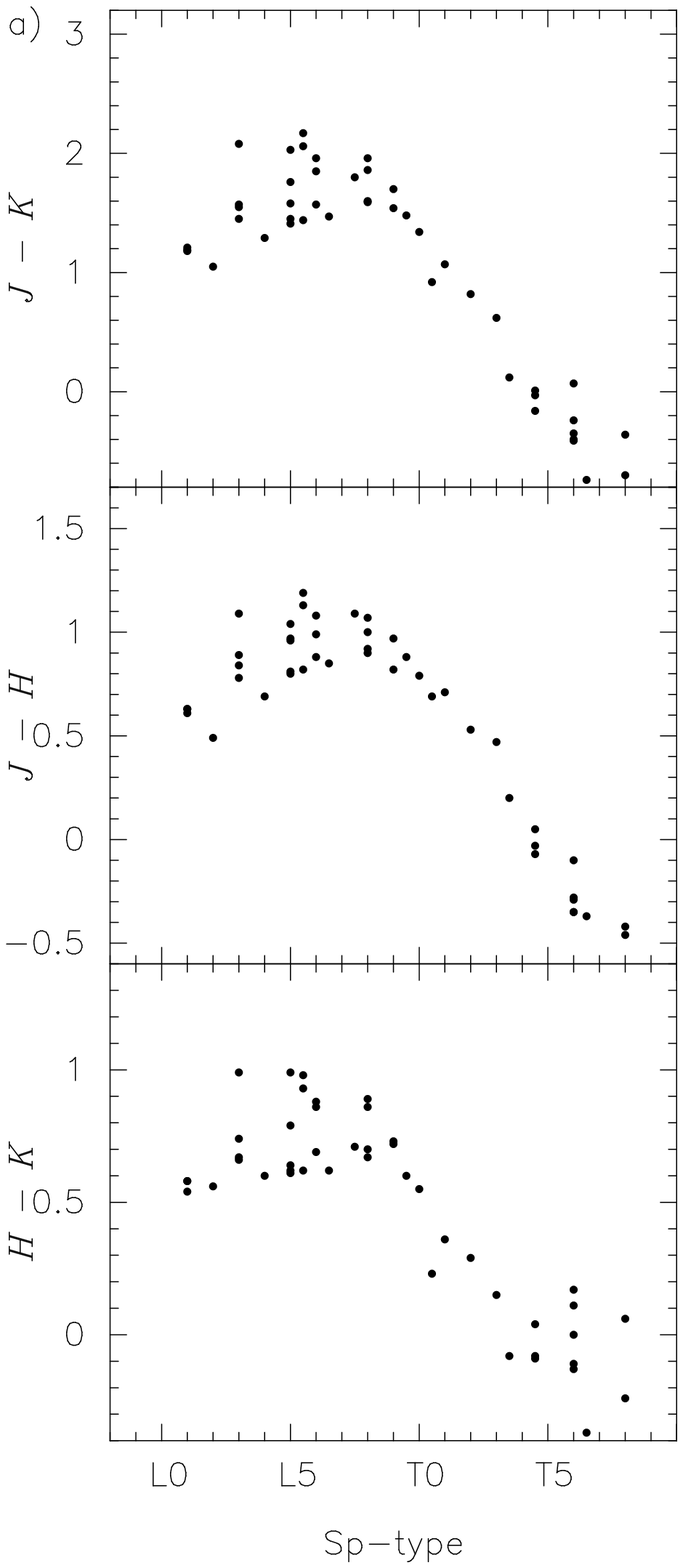}{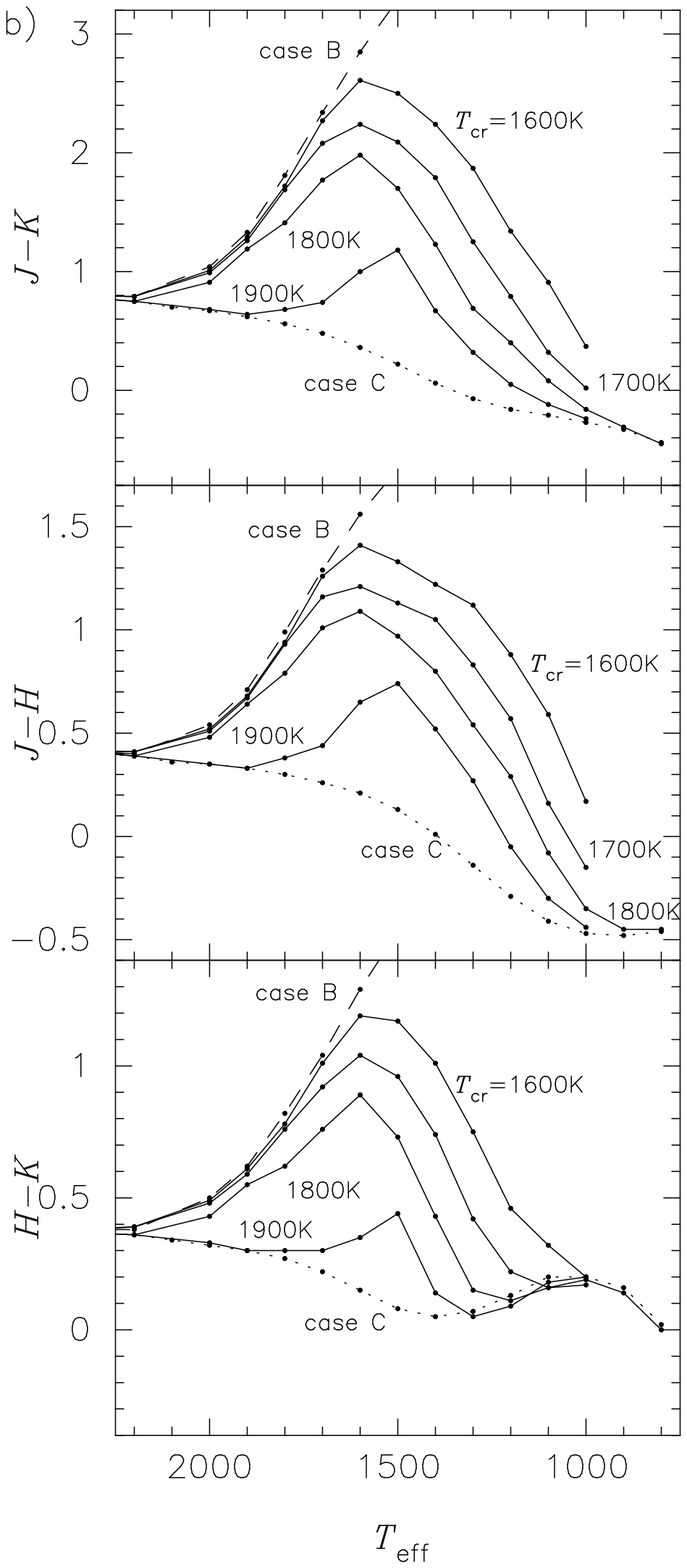}
\caption {
a) The observed  $J-K, J-H$, and $H-K$ colors (Leggett et al. 2002) are 
plotted against the spectral types (Geballe et al. 2002) in the top, 
middle, and bottom panels, respectively.
b) The predicted  $J-K, J-H$, and $H-K$ colors are plotted against 
$T_{\rm eff}$  in the top, middle, and bottom panels, respectively. 
The results based on the dusty models (case B), cloudy models of
$T_{\rm cr} = 1600, 1700, 1800,\& 1900$\,K, and dust-segregated models
(case C)  are shown by the dashed, solid, and dotted lines, respectively. 
}
\label{Fig8}
\end{figure}

\clearpage

\begin{figure}
\epsscale{0.75}
\plottwo{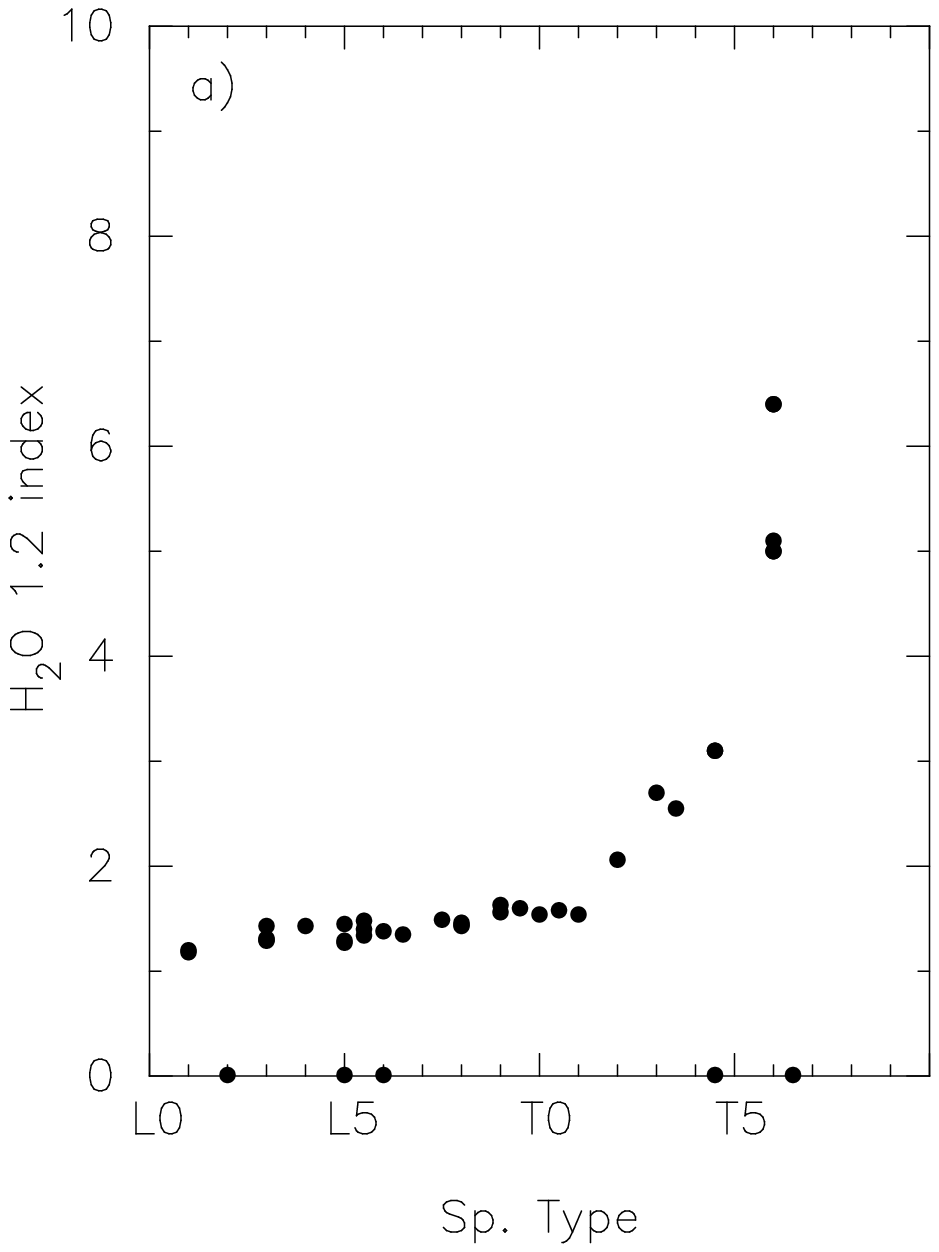}{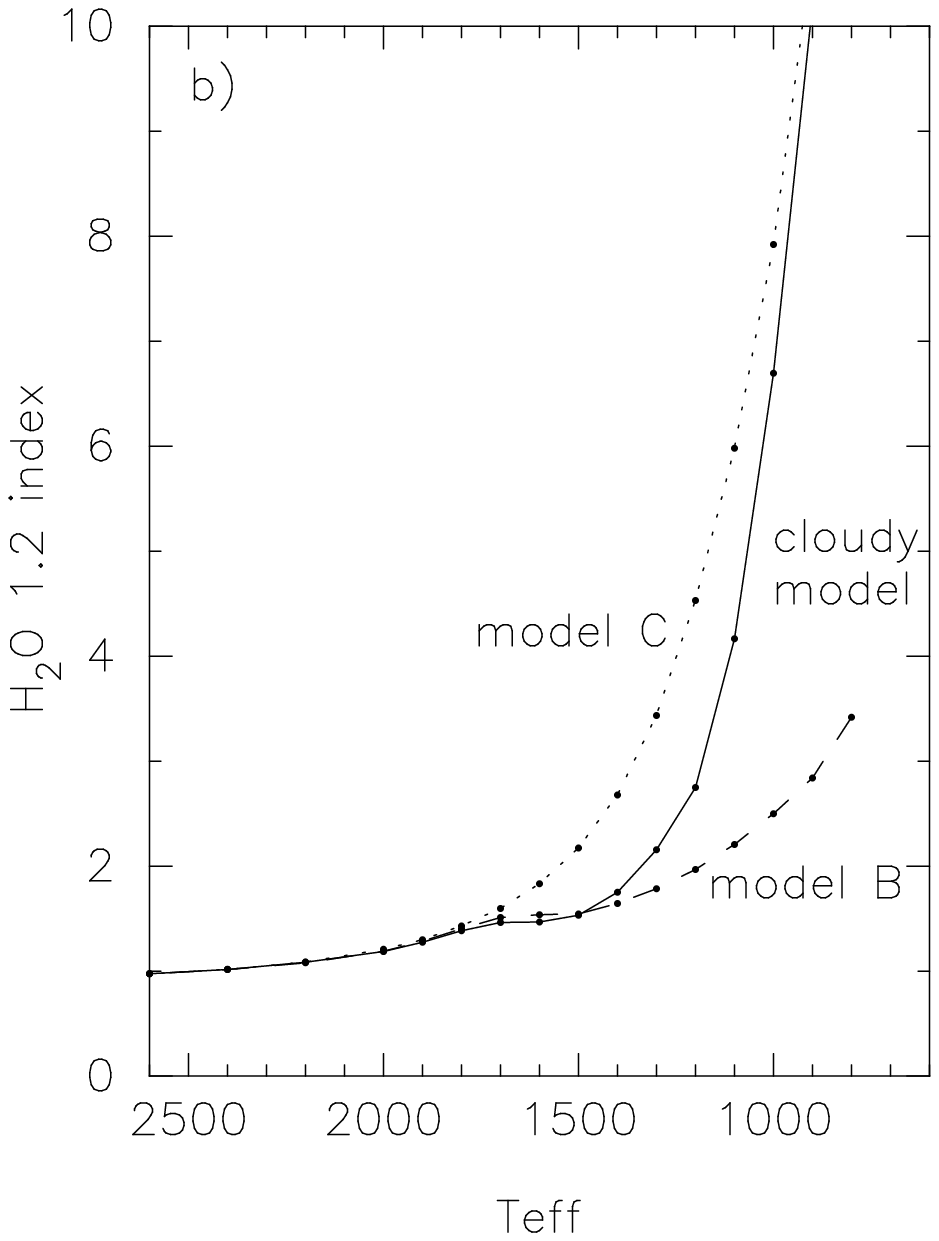}
\caption{
a) The observed H$_{2}$O 1.2 $\mu$m index defined as 
$F_{\lambda}(1.26-1.29\mu$m)/$F_{\lambda}(1.13-1.16\mu$m) by Geballe 
et al. (2002) is plotted against their Sp-type.
b) The predicted H$_{2}$O 1.2 $\mu$m indices based on the case B models,
cloudy models of  $T_{\rm cr} = 1800$\,K, and case C models are
plotted against $T_{\rm eff}$ by the dashed, solid, and dotted lines,
respectively.
}
\label{Fig9}
\end{figure}

\begin{figure}
\epsscale{0.75}
\plottwo{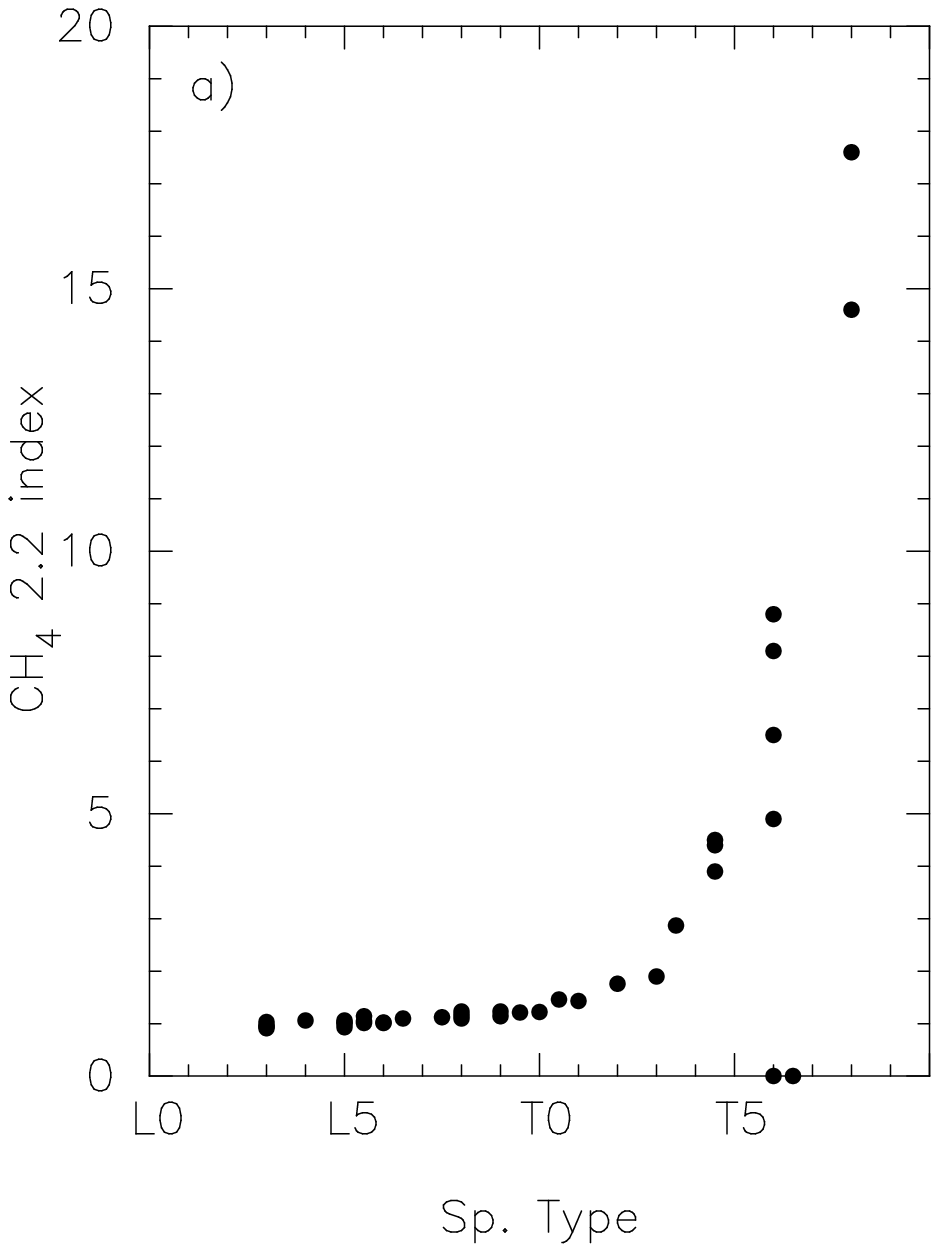}{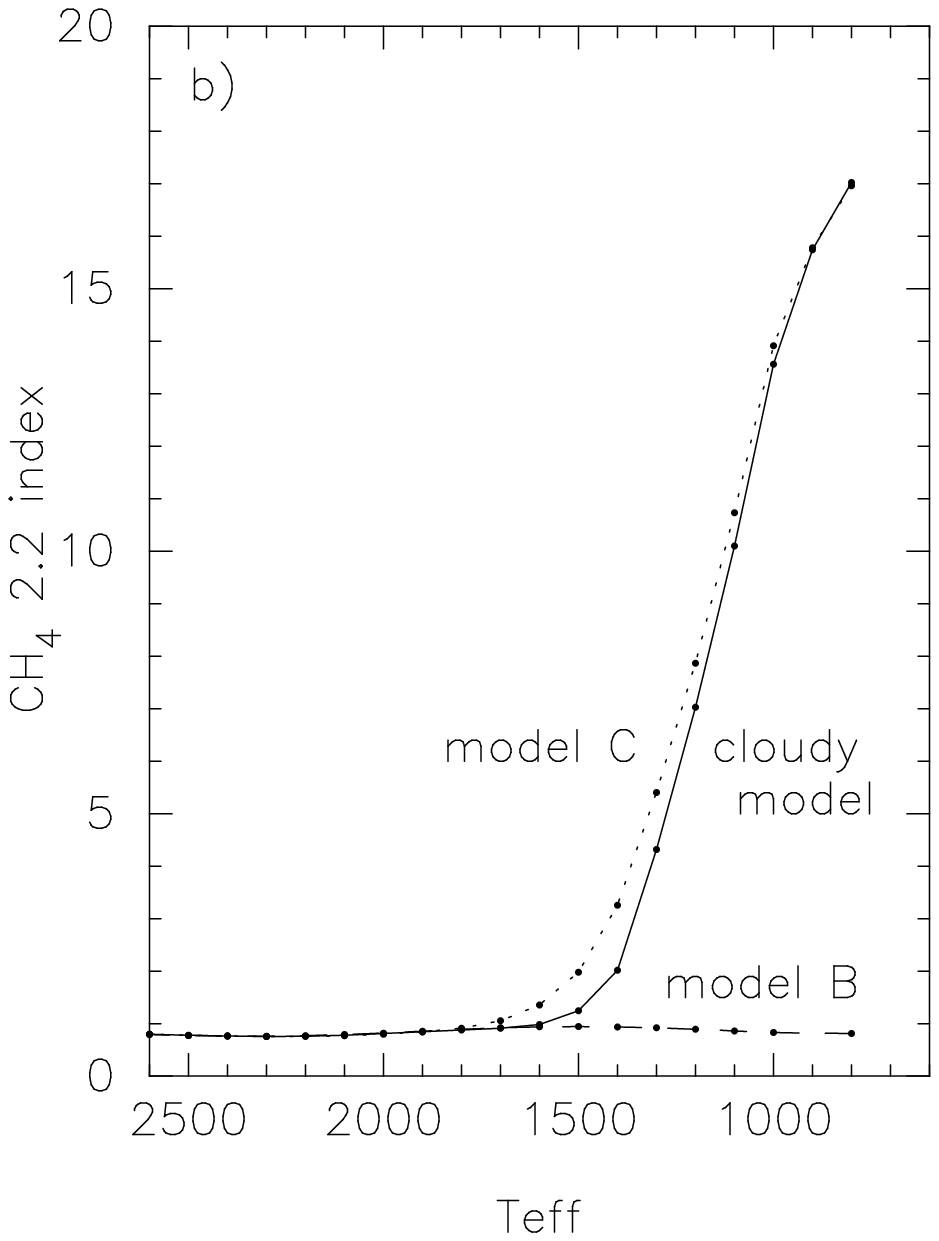}
\caption{
a) The observed CH$_{4}$ 2.2 $\mu$m index, $F_{\lambda}(1.26-1.29\mu$m)/
$F_{\lambda}(1.13-1.16\mu$m)  by Geballe et al. (2002) is plotted 
against their Sp-type.
b) The predicted CH$_{4}$ 2.2 $\mu$m indices based on the case B models,
cloudy models of  $T_{\rm cr} = 1800$\,K, and case C models are
plotted against $T_{\rm eff}$ by the dashed, solid, and dotted lines,
respectively.
}
\label{Fig10}
\end{figure}

\clearpage

\begin{figure}
\epsscale{0.75}
\plottwo{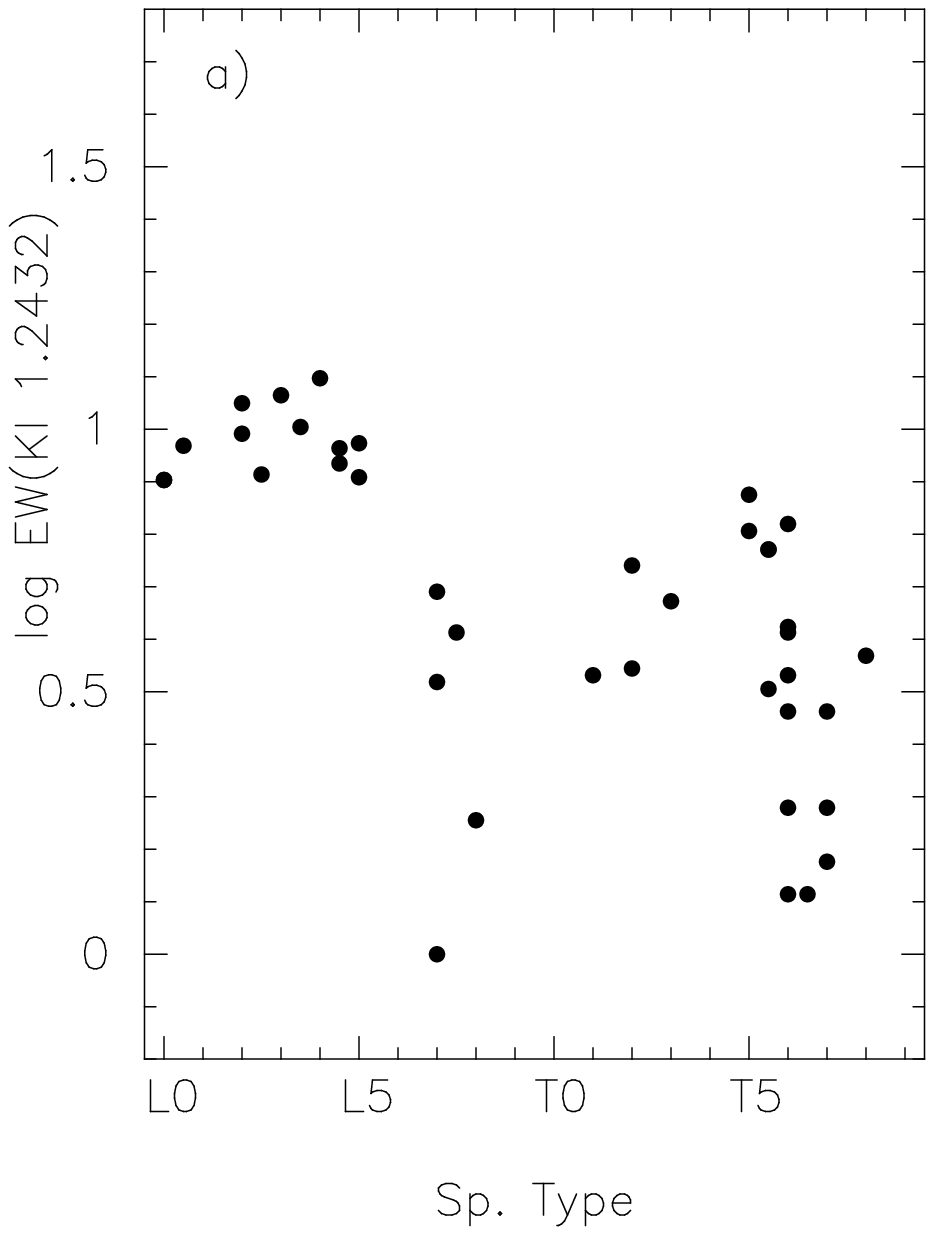}{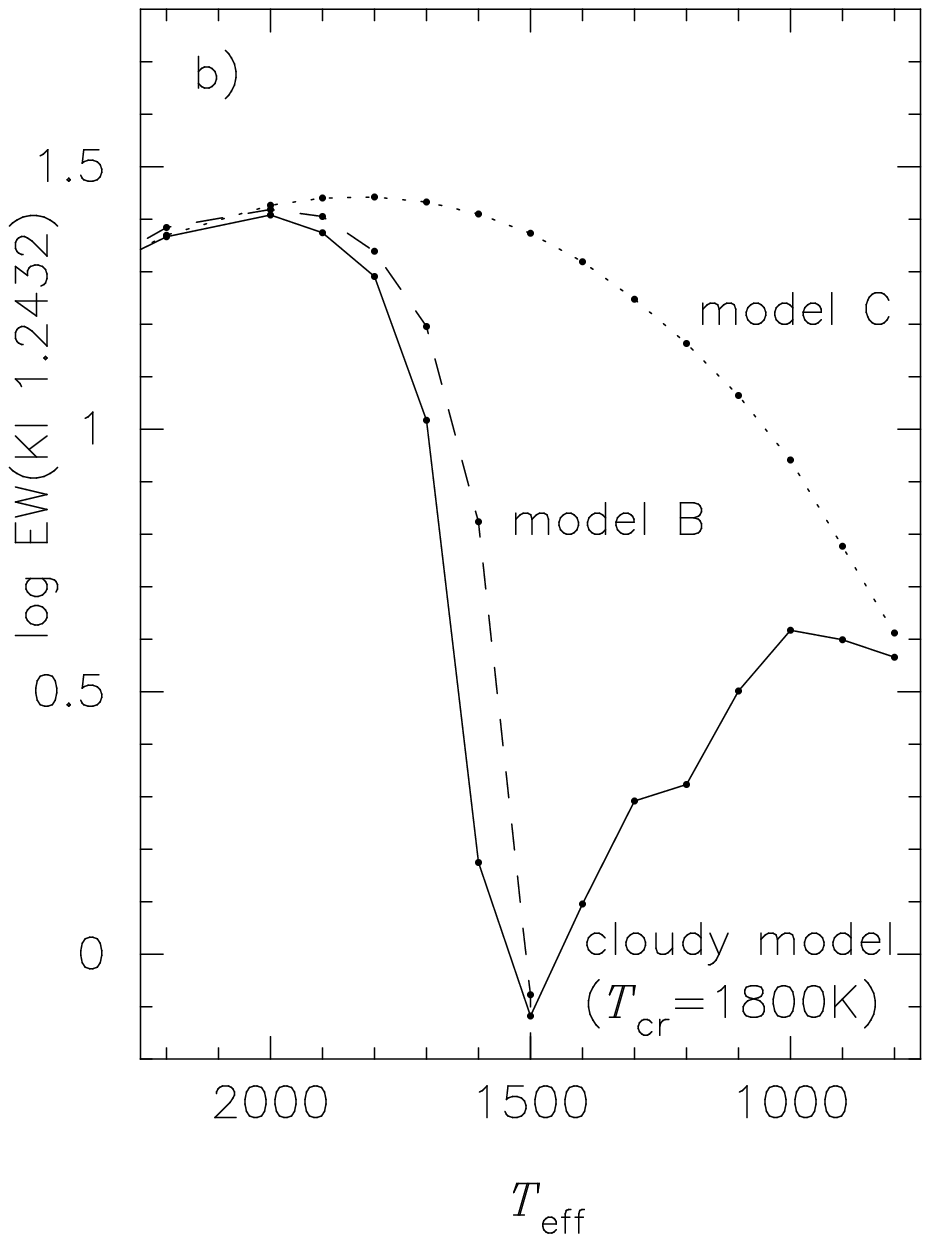}
\caption{
a) The observed  pseudo-EW's of K I 1.2432 $\mu$m given by Burgasser et al. 
(2002) are plotted against their Sp-type. 
b) The EW's of K I 1.2432 $\mu$m  measured on the predicted spectra based 
on the case B models, cloudy models of  $T_{\rm cr} = 1800$\,K, and case C 
models are plotted against $T_{\rm eff}$'s of the models by the dashed, 
solid, and dotted lines, respectively..
}
\label{Fig11}
\end{figure}

\clearpage

\begin{figure}
\epsscale{0.75}
\plotone{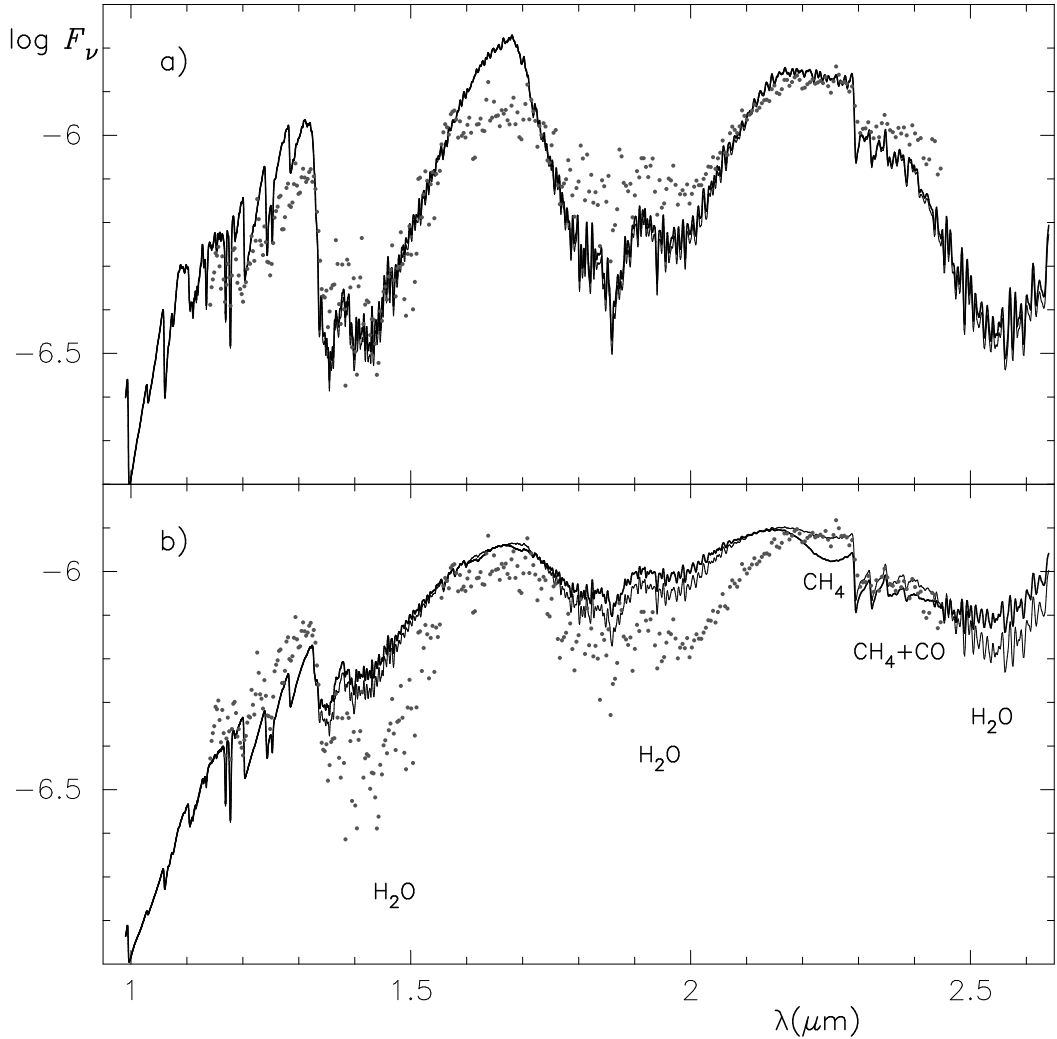}
\caption{
Comparison of the observed infrared spectrum of GD\,165B (Jones et al. 1994)
shown by the dots with the predicted ones based on the cloudy models of  
$T_{\rm cr} = 1800$\,K and $T_{\rm eff} = 1800$\,K, showing the effects of 
the oxygen abundance as well as of the silicate formation on the molecular 
abundances. a) The case of the high oxygen abundance (log $A_{\rm O} = 8.92$ 
as in the Table 1) is shown with and without the effect of silicate formation 
on the molecular abundances by the heavy and thin solid lines, respectively. 
b)  The same as in a) but for the case of the low oxygen abundance 
(log $A_{\rm O} = 8.69$ by Allende Prieto et al. 2001).
}
\label{Fig12}
\end{figure}

\clearpage

\begin{figure}
\epsscale{0.75}
\plotone{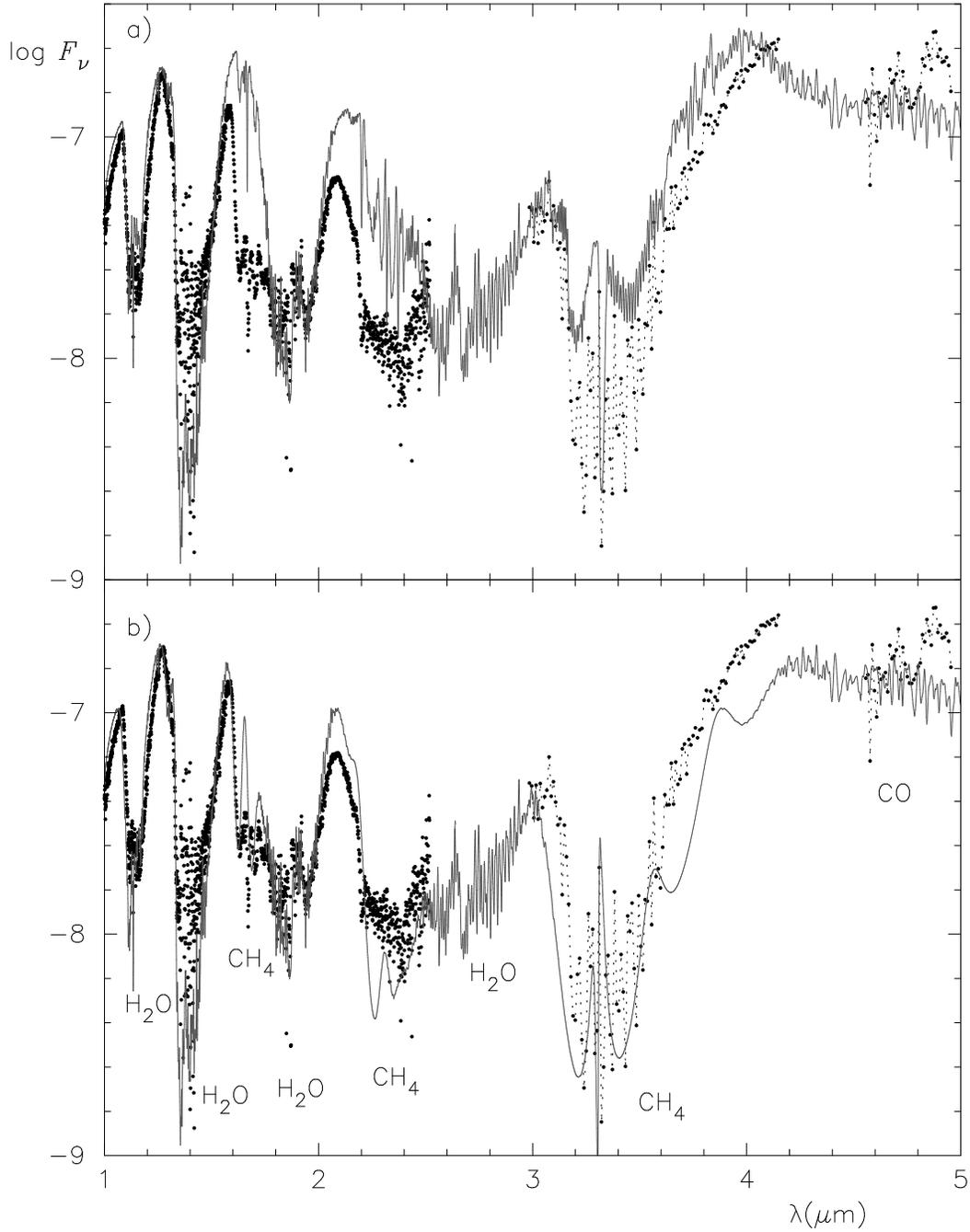}
\caption{
a) Comparison of the observed infrared spectrum of Gl\,229B (Geballe et al.
1996; Oppenheimer et al. 1998; Leggett et al. 1999) with the
predicted one using the GEISA linelist of methane on the cloudy model
of $T_{\rm eff} = 1000$\,K with  $T_{\rm cr} = 1800$\,K.
b) Comparison of the same observed  spectrum of Gl\,229B with the
predicted one using the band model opacity of methane on the same cloudy model
of $T_{\rm eff} = 1000$\,K with  $T_{\rm cr} = 1800$\,K.
The observed data and predicted spectra are shown by the dots and
grey lines, respectively.
}
\label{Fig13}
\end{figure}

\clearpage

\begin{figure}
\epsscale{0.75}
\plotone{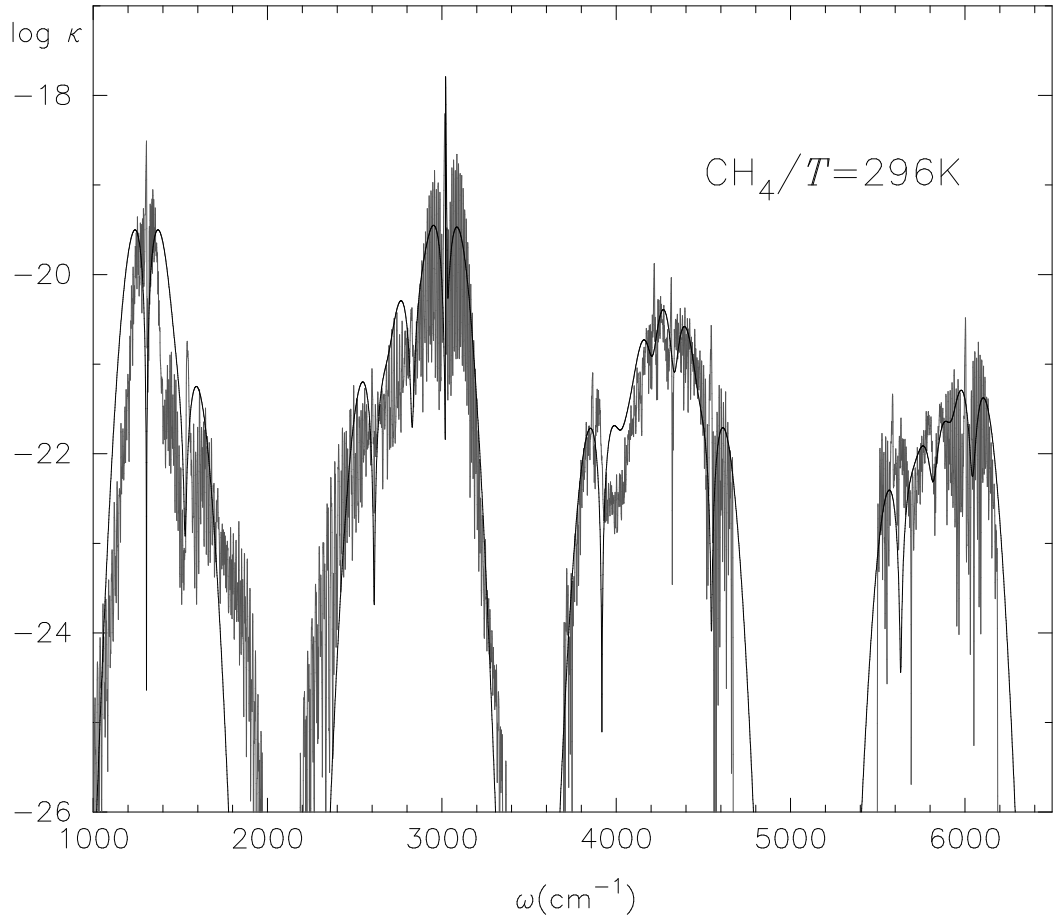}
\caption{
Absorption cross-sections (cm$^{2}$) of methane at $T = 296$K
evaluated with the use of the GEISA database and with the
band model are shown by the grey and black lines, respectively. 
}
\label{Fig14}
\end{figure}

\end{document}